 \documentclass[preprint]{aastex}

\slugcomment{}

\shorttitle{Djorgovski et al.}
\shortauthors{Collapsed Cores in Globular Clusters}

\begin{document}


\title{The Canada-UK Deep Submillimeter Survey: IV. The Survey
of the 14-Hour Field}


\author{Stephen Eales\altaffilmark{1}, Simon Lilly\altaffilmark{2},
Tracy Webb\altaffilmark{2}, Loretta Dunne\altaffilmark{1}, Walter
Gear\altaffilmark{1},David Clements\altaffilmark{1} 
and Min Yun\altaffilmark{3}}

\altaffiltext{1}{Department of Physics and Astronomy, Cardiff University,
P.O. Box 913, Cardiff CF2 3YB, UK}

\altaffiltext{2}{Department of Astronomy, University of Toronto, 60
St. George Street, Toronto, Ontario M5S 1A1, Canada}

\altaffiltext{3}{National Radio Astronomy Observatory, P.O. Box 0,
1003 Lopezville Road, Socorro, NM 87801}



\begin{abstract}

We have used SCUBA to survey
an area of $\rm \simeq 50\ arcmin^2$, detecting 19 sources
down to a 3$\sigma$ 
sensitivity limit of $\sim$3.5 mJy at 850$\mu$m. 
Monte-Carlo simulations 
have shown that the fluxes of sources in this and similar
SCUBA surveys are
biased upwards by the effects
of source confusion and noise, leading to an overestimate
by a factor of $\sim$1.4 in
the fraction of the 850$\mu$m background that has been
resolved by SCUBA. Once a correction is made for this
effect, about 20\% of the background has been resolved.
The simulations have also been used 
to quantify the effects of confusion on
source positions.
Of the 19 SCUBA sources, five are $\mu$Jy radio sources and
two are ISO 15$\mu$m sources. The radio/submillmetre flux
ratios imply that the dust in these
galaxies is being heated by young stars rather than AGN.
The upper limit to the average
450$\mu$m/850$\mu$m flux ratio implies either
that the SCUBA galaxies are at $\rm z >> 2$ or, if they are
at lower redshifts, that the dust is generally colder than 
in ULIRGs.

We have used 
simple evolution models to address the major questions
about the SCUBA sources: (1) what fraction of the star formation
at high redshift is hidden by dust? (2) Does the submillimetre
luminosity density reach a maximum at some redshift? (3) If the
SCUBA sources are proto-ellipticals, when exactly did ellipticals
form? We show, however, that the observations are not yet good
enough to answer these questions. There are, for example, acceptable
models in which 10 times as much high-redshift star formation is hidden by
dust as is seen at optical wavelengths, but also acceptable
ones in which the amount of hidden star formation is less than
that seen optically. There are also acceptable models in which very little
star formation occurred before a redshift of three (as
might be expected in models of hierarchical galaxy formation), 
but also ones
in which 30\% of the stars have formed by this redshift.
The key
to answering these questions are measurements of the
dust temperatures and redshifts of the SCUBA sources.

\end{abstract}




\section{Introduction}

The extragalactic background contains all the
energy ever emitted by galaxies, with the radiation from galaxies
at a redshift $z$ being weighted by a factor $(1+z)^{-1}$.
Thus
determing the level
of the background in different wavebands and resolving the background
into individual sources is of great importance for determing the
history of the energy output of galaxies.
Recent measurements suggest that (setting aside
the cosmic microwave-background radiation from the very early
universe) the
extragalactic background 
is dominated by emission in two spectral
regions, the optical/near-infrared
and the submillimetre wavebands ($\rm 100 \mu m < \lambda <
1 mm$), with roughly equal integrated emission
in the two wavebands \cite{dwek}. The spectral shape of the background
emission in the submillimetre waveband is characteristic of dust
emission spread over a range of redshift (Puget et al. 1996;
Lagache, Puget and Gispert 1999).
Therefore, although there is some
uncertainty from the $(1+z)^{-1}$ weighting factor,
this approximate equality
suggests that half the energy ever directly emitted by stars and active
galactic nuclei (AGN) has
been absorbed by dust and then re-radiated at long wavelengths.
Since young stars are generally more heavily obscured than old stars,
the fraction the of the universe's star formation that
is hidden by dust may well be significantly
greater than half.
The only caveat to these arguments is if there is some more exotic
population of objects dominating the submillimetre background
(Bond, Carr \& Hogan 1986, 1991).

If the extragalactic background is dominated by stars,
the history of the energy output of galaxies is closely linked
to the
history of star formation in the universe.
The background radiation and the local properties of galaxies
give complementary information about this history.
Consider the problem of the origin of elliptical galaxies and
spiral bulges.
The 
stars in nearby ellipticals and spiral bulges are often extremely old
(Bower, Lucey \& Ellis 1992),
implying that much of the star formation in these objects must have
occurred at very early times.
Ellipticals and spiral bulges in this initial star-forming phase
should be very luminous,
yet optical surveys have failed to
find convincing evidence of such ``proto-spheroids'' \cite{dep}.
However, these objects must make a major contribution to
the extragalactic background radiation.
The relation between the
integrated background emission produced by a population at
a redshift $z$ and the average cosmic density
of processed material, $<\rho (Z + \Delta Y) >$, produced by
this population is:
 
\medskip
$$
\int^{\infty}_0 I_{\nu} d\nu = {0.007 <\rho (Z + \Delta Y) > c^3 \over
4 \pi (1+z)}, \eqno(1)
$$
\medskip

\noindent \cite{pag}. Given the
quantity of metals associated with nearby ellipticals \cite{mike}, this
relation
implies that ellipticals in their rapid star-forming phase
should be responsible for about
half the background radiation. Thus resolving the extragalactic background
into individual sources {\it must} solve this problem.
 
The measurement of the integrated optical background
recently made by Bernstein and collaborators \cite{reb}
is only about twice that obtained by simply adding up the emission
in deep galaxy counts
\cite{hdf} and indeed Bernstein argues that
the integrated optical background can be explained by
known populations of objects.
Thus the optical background may already have
been completely resolved and no obvious proto-spheroids have been
found. For example, the galaxies at $\rm z \sim 3$ found by
the Lyman-break technique \cite{chuck} contribute $\sim$2\%
of the background in the I-band and thus, by the metallicity argument above,
are unlikely to be be the missing proto-spheroids. This apparent
failure
to find proto-spheroids at optical wavelengths, together with the
metallicity argument, suggests that proto-spheroids must
dominate the other important component of the background at
submillimetre wavelengths. A physical explanation of this would be
if the initial formation of stars in the spheroids leads to the rapid
creation of dust, which then absorbs the optical/UV radiation 
\cite{ee96,ee97}.
 
Resolving the submillimetre
background radiation into individual sources
is thus of great interest. It has recently become possible
to do this, at least partially,
with the
ISOPHOT instrument on the Infrared Space Observatory at short
wavelengths and 
with the SCUBA submillimeter array \cite{wayne}
on the James Clerk
Maxwell Telescope (JCMT) at long wavelengths. This 
paper is the fourth of a series of papers
describing the results of a deep submillimetre survey with SCUBA
(Eales et al. 1999; Lilly et al. 1999---henceforth Papers I and II).
Before describing the scope of the present paper,
we will briefly summarize the results of our
and other submillimetre surveys.

The ISOPHOT surveys are important because the
submillimetre background radiation peaks at short wavelengths, and
thus, in energy terms, it is more important to resolve the background
here than at the longer wavelengths sampled by SCUBA.
Surveys
with ISOPHOT have found sources that contribute about
10\% of the submillimetre background at 175 $\mu$m \cite{kawara,puget2}.
However, the large size of the ISOPHOT beam 
(1.9 arcmin, full-width half-maximum)
means that it is remarkably difficult to identify
the galaxies responsible for the submillimetre emission
\cite{scott}.
 
The main operating wavelength of SCUBA is 850$\mu$m, where the
submillimetre background is much lower than at its peak ($\nu I_{\nu}$
lower by a factor of $\sim$30), but SCUBA has the great advantage over ISOPHOT
that its much smaller beamsize (14 arcsec, full-width half-maximum)
means that it is possible (although still difficult) to identify
the galaxies responsible for the submillimetre emission.
SCUBA has been used to investigate the high-redshift universe
through three different methods. The first is to
obtain submillimetre images in the directions of rich clusters,
thus using the lensing effect of the clusters both to amplify the
submillimetre fluxes of high-redshift submillimetre sources and
to reduce the effect of source confusion 
(Smail, Ivison \& Blain 1997; Smail et al.
1998; Blain et al. 1999b).
The disadvantages of this are the need to separate
background from cluster sources and for
an accurate lensing model for each cluster.
The second method is to carry out truly `blind-field' submillimetre
surveys, which do not have the disadvantages of the cluster method
but at the expense of greater problems with confusion and much
longer integration times to get to the same effective sensitivity.
The deepest blind survey that has so far been carried out
is a survey of the
Hubble Deep Field \cite{dave}.
Our survey, which is not so deep but which covers a wider
area, is also of fields which have been extensively surveyed
at other wavelengths: the Canada-France Redshift Survey fields
(CFRS; Lilly et al. 1995). 
The Hawaii group has been carrying out a survey with a similar depth
to our own
(Barger et al. 1998; Barger, Cowie \& Sanders 1999). 
The third of the methods is to carry out SCUBA observations
of known high-redshift objects. This will be discussed below but
we will first summarize
the scientific
results of the surveys.

It has been claimed (Blain et al. 1999b,c) that the SCUBA
surveys have resolved close to 100\%
of the background at 850$\mu$m,
although we will argue in this paper that this is an
overestimate.
Despite controversy over the details of which galaxies are responsible
for the submillimetre emission, caused by the large
positional errors of SCUBA, all
the teams carrying out SCUBA surveys agree that these galaxies are similar to
the ultra-luminous infrared galaxies (ULIRG's; Sanders
\& Mirabel 1996)
found
in the local
universe \cite{dave, barg, pap2, smail2}.
From these two results one
can draw an extremely important
conclusion (Paper I). If ULIRG's make up $>$50\% of the
submillimetre background, they must constitute $>\simeq$25\%
of the total extragalactic background radiation; and thus, as long as
the dust in ULIRG's is heated by stars rather than AGN,
roughly one quarter of all the stars that have ever formed
must have formed in extreme systems like this rather than in normal
galaxies like our own. This fact, together with the extreme
bolometric luminosities of ULIRG's, which imply star-formation
rates of $\rm 10^2-10^3\ M_{\odot}\ year^{-1}$ \cite{rieke}, 
make it difficult to avoid the conclusion
that these systems are proto-spheroids. 

Hughes et al. (1998) and Barger, Cowie and Richards
(2000) have also argued that the
results of the SCUBA surveys and the strength of the submillimetre
background imply the amount
of star formation hidden by dust in the early universe
is an order of magnitude greater than that seen
directly in optical surveys---in effect that the
universe had a `Dark Age' in which the fraction of star
formation hidden by dust was much greater than it
is today.
In our earlier work, we found little evidence
for this, concluding that the cosmic evolution seen in the
submillimetre waveband is similar to that seen at optical
wavelengths and thus that the fraction of young stars
hidden by dust is the same at all cosmic epochs (Paper I).
Moreover, the fraction of the SCUBA sources identified with galaxies
at $\rm z < 1$ suggested that the submillimetre luminosity-density
declines at $\rm z > 2$ (Paper II),
which is where the other studies
claim the majority of the dust emission is occurring.
We will revisit this question in this paper.

There are two fundamental uncertainties in the conclusions
above.
The first of these is simply the sensitivity of these
conclusions to dust temperature, which for the SCUBA
sources is largely unknown. Up to $\rm z \sim 3$, the main
SCUBA operating wavelength of 850$\mu$m falls, in the
rest frame, on the Rayleigh-Jeans
side of the typical dust spectral energy distribution, and thus
dust temperature is critical for calculating the total dust luminosity;
for a dust source with a
dust emissivity index of two, the total dust emission
is proportional to the dust temperature to the sixth power, and
thus an uncertainty of a factor of two in dust temperature
leads to an uncertainty of a factor of 64 in total luminosity.
This uncertainty holds for both individual sources and for the
population of SCUBA sources as a whole. Even the conclusion that
a SCUBA galaxy is essentially a ULIRG is affected. Fig. 1 shows
the bolometric luminosity calculated for a SCUBA source with
$\rm S_{850 \mu m} = 4 mJy$ as a function
of redshift and for a number of assumed dust temperatures.
Also shown is the bolometric luminosity of the archetypical ULIRG
Arp 220. If instead of assuming a dust temperature typical
of a ULIRG, for which there is rarely any observational
justification, one assumes
a dust temperature typical of a normal spiral galaxy, one obtains
a bolometric luminosity an order-of-magnitude lower than that of Arp 220.
Thus, the argument that the estimated bolometric luminosities
imply SCUBA galaxies are ULIRGs is 
circular: if one
assumes the dust temperature typical of a ULIRG, one necessarily
obtains the
bolometric
luminosity of a ULIRG. We will explore the importance
of this uncertainty for the population as a whole later
in this paper.

The second of the uncertainties is whether the dust in a SCUBA
galaxy is actually being heated by young stars or whether
it is being heated by an obscured active galactic nucleus (AGN);
if it is the latter, of course, then the conclusion that the SCUBA
galaxies are proto-ellipticals is completely wrong.
Two results suggest that the AGN hypothesis may be the correct one.
First, the
recent discovery that there are black holes in the centres
of most nearby galaxies \cite{magor}
suggests that most galaxies have passed through
a phase in which they harboured an active nucleus, implying in turn
that AGN might contribute a significant fraction
of the extragalactic background radiation. Estimating the exact fraction
of the total extragalactic background emission that is produced by AGN
is difficult, mainly because one does not know how efficiently mass was
turned into radiation as the black hole formed (Haehnelt, Natarajan
and  Rees 1998), and even simple estimates
range from 10\%
\cite{pap1} to 60\% (Lilly, unpublished). Second, it has  been known
for a long time \cite{setti} that one way of explaining the spectral dependence
of the X-ray background is by a new population of highly obscured AGN,
and the submillimetre waveband would be the natural place for the
absorbed emission to reappear.

Settling this issue is unfortunately difficult.
Although the evidence from mid-infrared line ratios \cite{genzel,lutz}
and from VLBI radio observations \cite{smith}
is that stars rather than AGN
are the main energy source in low-redshift ULIRGs, it
is not possible to use these techniques for the SCUBA galaxies.
One result that does suggest that this is also the
case for the SCUBA galaxies is that
they are often detected in very deep radio surveys with
ratios of dust to radio emission similar to those
seen in samples of nearby star-forming galaxies
(Lilly et al. 1999; Barger, Cowie \& Richards 2000).
Barger et al., for example, found that five of the seven sources
detected in a SCUBA survey  of the Hubble Flanking Fields
were detected in the radio survey of the same region.
One way of addressing the possibility that the SCUBA galaxies
are the obscured AGN needed to explain the X-ray background
would be deep X-ray surveys of fields surveyed with SCUBA, something
now possible with Chandra and Newton. The first of these surveys
\cite{fabian,horn} suggest that it is young stars rather than AGN 
heating the dust.

The third of the methods used to investigate the high-redshift
universe with SCUBA has been to use SCUBA to observe known classes of object.
Observations of samples of high-redshift radio galaxies have shown
that the average submillimetre luminosity of these objects rise
as $(1 + z)^3$ out to $\rm z \sim 4$ \cite{arch}, very different
from the ways in which the luminosity-density of the galaxy 
population as a whole is inferred to evolve, either from
optical \cite{chuck} or submillimetre \cite{blain,pap1,pap2}
observations. SCUBA observations have also been carried
out of the high-redshift galaxies found using the
Lyman-break technique \cite{chuck}. Observations of individual
galaxies have resulted in only a single detection \cite{chap},
but Peacock et al. (2000) have claimed a statistical detection
of the population by summing the submillimetre emission at the
positions of Lyman-break galaxies in the deep SCUBA image
of the Hubble Deep Field. This has implications for
whether the Lyman-break galaxies and
the galaxies found in the deep SCUBA surveys are distinct populations.
They can not be completely separate, of course, because any SCUBA
galaxy at the right redshift will necessarily have a Lyman break
in its $UV$ spectrum;
the question is really whether, as Adelberger
and Steidel (2000) have recently claimed, one can learn all one needs to know
about high-redshift galaxies by studying the Lyman-break
population and making the necessary bolometric corrections for each galaxy
to allow for the dust emission. However, the optical/IR
observations of the SCUBA sources suggest that this is not correct, because
even the brightest SCUBA sources frequently have optical counterparts
that are too faint to have been detected in any of the optical searches
for Lyman-break galaxies (e.g. Gear et al. 2000).
We 
will investigate
this question in more detail later in this paper.

In Paper I of this series (Eales et al. 1999) 
we described submillimetre observations of three fields:
two small fields within the CFRS 3$^h$ and 10$^h$ fields and a larger
field within the CFRS 14$^h$ field. Paper II (Lilly et al. 1999)
considered the optical,
radio and mid-infrared properties of these fields in an attempt
to identify and determine the properties of the
galaxies responsible for the submillimetre emission.
Paper III (Gear et al. 2000) describes interferometry at
millimetre wavelengths and deep optical/IR imaging and
spectroscopy of the
brightest SCUBA source in the 14$^h$ field.
After our initial observations of the two small fields, our strategy
changed to a systematic attempt to map at submillimetre
wavelengths as great a fraction as possible
of the CFRS 3$^h$ and 14$^h$ fields, which are the
two with ISO mid-infrared images. In this paper we report
the results of our submillimetre survey of the 14$^h$ field and
compare the submillimetre, radio and mid-infrared results, in an attempt
to understand the properties of the SCUBA galaxy population. Subsequent
papers will describe the optical and near-infrared properties of this
field and also the results of the survey of the 3$^h$ field.
 
In this paper we will use the results of another
survey, the SCUBA Local Universe Galaxy
Survey (SLUGS; Dunne et al. 2000). This is a parallel submillimetre
survey of the
local universe meant to serve as a zero-redshift benchmark
against which the surveys of the high-redshift universe can be compared.
Dunne et al. have recently published the first estimate of
the local 850$\mu$m luminosity function based on SCUBA
observations of a sample of 104 galaxies
selected from the IRAS Bright Galaxy Sample \cite{soifer}.
We will use this luminosity function and a number of
other results from SLUGS later in this paper. 
 
The arrangement of this paper is as follows. Sections 2 and 3
describe the
observations and the data reduction. Section 4 describes
how a catalog of sources was selected from the final image and
includes
a detailed investigation of the reliability of the catalog which
is applicable to all SCUBA surveys.
Section 5 describes
a search for associations of SCUBA sources with mid-infrared and
radio sources. Section 6 contains a discussion
of 
the results of the current survey and 
includes an estimate of the redshift distribution.
Section 7 describes detailed modelling of the SCUBA galaxy population
and a discussion of its cosmological significance. 
We everywhere assume a Hubble Constant of
$\rm 75\ km\ s^{-1}\ Mpc^{-1}$.

\section{Observations}

We observed the 14$^h$ field with SCUBA
on 20 nights between 1998 March 5
and 1999 May 27. SCUBA is described in detail elsewhere \cite{wayne} but,
briefly, it consists of 91 bolometers for observations at short
wavelengths, usually 450 $\mu$m, and 37 bolometers for observations
at long wavelengths, usually 850 $\mu$m; a dichroic beamsplitter
is used 
to simultaneously observe the same field at the two wavelengths.
The field-of-view of the array is roughly circular with a diameter
of about 2.3 arcmin. The beam size is about 8 arcsec and
14 arcsec (full-width half-maximum) at 450 and 850 $\mu$m. 
As the bolometers in the array do not fully-sample the sky, the
secondary mirror is moved in a hexagonal
pattern, producing a fully-sampled image known as a `jiggle
map' \cite{wayne}. SCUBA sits at the Naysmith focus of the JCMT
(i.e. fixed relative to the Earth)
so each bolometer gradually moves in a circular path around the field
centre.

For our observations we used a 64-point jiggle pattern,
which produces fully-sampled maps at both
wavelengths. As in classical infrared and submillimetre
astronomy, we chopped and nodded 
the JCMT secondary mirror to remove temporal fluctuations and linear
spatial gradients in the sky brightness, using
a `chop throw' of 
30 arcsec in Right Ascension.
This small angle, only twice the size of the JCMT beam,
maximizes the accuracy of the sky removal
and also means that a source is likely
to remain on the array even when the array is centered at the
reference position, increasing the effective
integration time on the source. This does, however, introduce the
complication that any real source will appear
as a positive peak with negative peaks of half the
amplitude 30 arcsec to the east and west of the positive peak.
The individual units of our survey were jiggle-map observations at
a number of field centres within the overall survey region.
Each observation lasted about one hour.
The field centres were chosen with the aim of producing a map with
uniform noise.
Each point in the final map
incorporates data from a large number of separate
datasets (typically nine) and from a large number of
different bolometers, since the bolometers
move with respect to the sky.
This lessens the chance of problems with individual
bolometers generating spurious sources.

Before each individual observation we checked the pointing of
the telescope on a nearby
JCMT pointing source. 
During each night we monitored the opacity
of the atmosphere at both 450 and 850 $\mu$m using `skydips' and
we determined the flux calibration from observations
of one or more of the following calibrators: Mars, Uranus, 
CRL618, IRC$+$10216. The photometric accuracy of SCUBA
observations at 850 $\mu$m is remarkably good and we estimate that
the basic photometric uncertainty (i.e. the flux error
for an object with
infinite signal-to-noise) is only $\simeq$5\%; at 450 $\mu$m it is rather
worse, $\simeq$20\%, because the flux calibration
is much more sensitive at this wavelength 
to errors in the dish surface, which
depend strongly on the recent thermal history of the dish.

Details of the nights on which we observed are given in Table 1.




\section{Data Reduction}

We reduced the data independently in Cardiff
and Toronto using the SURF package \cite{tim} in
the following way.
First, for each bolometer we subtracted the intensity measured
at the reference position from the intensity measured
at the target position, dividing the result by
the array's flat-field to correct for sensitivity variations
between bolometers. We then corrected each second of data
for each bolometer
for the
atmospheric opacity at that time.
After these standard steps we then inspected intensity versus time plots
for each bolometer and excised obviously bad data, also running several
SURF clipping routines which automatically flag bad data.
Because of
the jiggling technique, each point on the final map is made from data
taken at a slightly different time, and thus a varying sky level can
lead to increased noise on the final map.
To avoid this, we ran the SURF program REMSKY, which subtracts from the
intensity measured by each bolometer the median of intensities
measured by all the bolometers at that time.
Even
in the relatively good conditions in which most of the data were taken, this
step always made a large improvement to the final maps.
Finally, we made a map, incorporating the data for each bolometer
in each survey unit with the optimum statistical weighting, and 
using a spatial linear
weighting function to regrid the data onto a rectangular mesh.

The final Cardiff 850 $\mu$m map is shown in Fig. 2 (the
Toronto version is very similar; the 450 $\mu$m map,
which has much poorer sensitivity, will be considered later).
An edge region containing artefacts,
a well-known phenomenom of the SCUBA map-making process,
has been removed from the
map.
The negative lobes of many of the sources can clearly be seen
and in the case of the brightest source (bottom right-hand quadrant)
the effects of an error in the telescope software can even be
seen in the slight offset of the lobes from an East-West
direction (Until August 1998
the telescope software was correctly translating the telescope
chop direction
into the telescope's natural altitude-azimuth coordinates at the
beginning of each integration but was not updating this
calculation during the integration, resulting in this small
rotation. The detailed effect of this on each source
is impractical to model, but our simple models show 
that the size of the effect is small
enough that an assumption of no rotation leads
to a negligible effect on the source catalog.).

We tried to improve our map in two ways. First, 
we tried a more sophisticated sky-removal algorithm than
that used in the REMSKY program, in which
each bolometer is assumed to see the same
sky brightness, by instead assuming that the sky brightnes is a linear
function of position. This algorithm
made a significant improvement to SCUBA data taken for other
projects in poor atmospheric conditions \cite{loretta} but
made negligible improvement to the map in Fig. 2. 
Second, we carried out a Fourier Transform of the data
for each individual bolometer to look for bad data.
This proved to be quite instructive, since although most
bolometers had white-noise spectra, some had spectra
in which the power was much greater at high frequencies;
and on re-examination of the data back in the time domain, these
always proved to be bolometers where there was some sign that
the signal was correlated with that in other bolometers
(possibly due to inadequate shielding between the electronics
associated with each bolometer). We tried drastic filtering
of these bolometers in the frequency domain. However, when the
data were transformed back into the time domain and a new map made, it
was very similar to that in Fig. 2.
This failure to improve the basic map does suggest, however,
that its features are robust
and are not the result of 
gradients in sky emission or of noisy bolometers. 

\section{The Catalog}

\subsection{Selecting the Sources}

To generate a catalog there are three problems to overcome:
how to use the information in the negative as well as the positive
lobes of each source; how to determine the significance of the
sources; how to combat the effects of source confusion.

As in Paper I, we solved the first
problem 
by convolving the
raw map with a template of what a real source should
look like in the absence of noise \cite{pd}. We used Uranus as the template.
We averaged all the maps of Uranus, which we
observed many times during the survey with the same chop
throw as for the survey, to provide a master template
and convolved this with the raw survey image. 
This
implicitly makes the assumption that a real
SCUBA source will not be significantly
extended, but this is a reasonable assumption
given the relative sizes of the SCUBA beam and of the
optical structures of high-redshift galaxies.

To model the noise, vital for determining the significance
of the sources,
we started with the basic assumption that the noise
on a bolometer 
is
independent of the noise on every other bolometer. We then took
each individual pointing of one hour duration
(the survey unit), measured the standard deviation of the intensities
for each bolometer, and replaced the real data with the output of
a Gaussian random-number generator with the same standard deviation
as the data. The real data does not have a precisely Gaussian form,
mainly because of the effects of some of the SURF routines, principally
the clipping routines and REMSKY. So we ran the same SURF programs
on the artificial data, and then rescaled the artificial data so
that it had again the same standard deviation as the real data. In this
way the histogram of intensities for an artificial bolometer is almost
identical to that for the real data. We did the same for each
bolometer within each survey unit, and then made a map with the
artificial datasets, repeating this whole procedure
until we had 1000 artificial maps. 
Since our source-selection procedure uses
the real map convolved with the template, we convolved the artifical
maps with the same template.
Fig. 2 shows the noise map produced
by measuring the standard deviation of these convolved maps, pixel by pixel. 
As one would expect, the noise is a strong
function of position. In particular, the noise in the upper third of
the map is significantly worse than the noise elsewhere, because 
part of the image was made with data taken in May 1999, during which the
atmospheric conditions were significantly worse than for the rest of the 
survey.
The noise estimated from the deep
area of the noise map is
0.94 mJy, whereas the noise on the real map after it has
been convolved with the template and after all
the significant sources have been subtracted (see below) is 0.90 mJy.
These are remarkably similar, especially as the real map should
contain a noise component due to faint sources below the sensitivity
limit of our catalog. 

The starting point for the catalog selection was the raw image
convolved with the template and then divided by the noise image.
This image is a map of signal-to-noise over the field. To address
the problem of confusion we used the CLEAN algorithm. We first
used the signal-to-noise map to produce a list of possible sources.
We then iteratively CLEANed the {\it raw} map in boxes centred
on the positions
of these possible sources. Then, for each source in the list
of possible sources, we adopted the
following procedure. First, use the information from CLEAN to
remove all other possible sources from the raw map. Then
convolve this new map with the source template and divide
the convolved map by the noise map, and so measure the signal-to-noise
of the possible source. In this way, we generated a catalog
of sources, producing (as in Paper I) fluxes and positions
from the template-convolved map, the optimum procedure for 
measuring fluxes and positions \cite{pd}.

An important question is the minimum signal-to-noise a source
should have to be included in the catalog. As in Paper I, we adopted
the pragmatic solution of carrying out exactly the same selection
procedure on the negative of the map, to determine the number of spurious
sources as a function of signal-to-noise. We adopted a minimum 
signal-to-noise of three, since this is the point at which the ratio
of spurious to real sources starts to climb sharply.

We independently carried out this procedure on the Cardiff and Toronto
maps and Table 2 contains the sources that met the 3$\sigma$ threshold in
both catalogs. The positions and fluxes are the averages of the values
for the two catalogs. There are 19 sources in the catalogue. Undoubtedly
a small number are spurious. We will address this whole problem in
more detail in the next section, but a simple argument suggests 
that 2-3 are probably false detections. First, applying the selection
procedure to the negative map resulted in two detections, both
at $\simeq$3$\sigma$. Second, 
given the number of beam areas in
our survey area, Guassian statistics imply that 
a 3$\sigma$ cutoff should produce 2.6 spurious
sources. 

Of the seven sources in this field listed in Paper I, three now fall
outside the sample. One of these, CFRS 14G, was right at the edge
of the region we thought free of edge artefacts but is clearly
such an artefact. The other two are
still detected (their new fluxes and positions 
are given in Table 3) but at a level below the nominal catalog
limit. This loss of sources is of course expected.
If sources are close to the signal-to-noise limit of a sample, a
small amount of new
data will result in approximately as many sources being lost
from the catalog as are gained.  

We measured 450 $\mu$m fluxes for the 850 $\mu$m sources by
aperture photometry with 
an aperture
of diameter 12 arcsec centered on the 850 $\mu$m position, using
the technique of Dunne et al. (2000) to obtain
the errors. Only one source is detected at $>$2$\sigma$ 
(Table 2).
It is possible that even if individual sources are not
detected at a significant level, the population as
a whole is. We tested this by
making a weighted average of the 450$\mu$m flux densities
in Table 2 (excluding the two most inaccurate values
which are for objects right at the edge of the slightly
smaller 450$\mu$m map and so are
probably unreliable). We obtained
$\rm <S_{450 \mu m}>=2.2 \pm 2.7\ mJy$ and thus did not obtain
a significant detection. The corresponding weighted
average at 850$\mu$m is $\rm <S_{850 \mu m}>=4.34 \pm 0.46\ mJy$.
This result will be discussed further in \S 6.

Fig. 3 shows the raw source counts with
a correction made for the sensitivity variation
over the image but none made for the
effect of flux errors. These agree
fairly well
with those of other surveys (Hughes et al.
1998; Barger, Cowie \& Sanders 1999; Blain
et al. 1999b). We examine the effects of incompleteness and
confusion on the sources counts in \S 4.3.

\subsection{Notes on Individual Sources}

In this section we give notes on individual sources.
The positional
accuracy of the SCUBA sources is important for determining
the optical/IR counterparts. Since we independently reduced
the data and compiled catalogs in Cardiff and Toronto,
one partial check on the accuracy is whether the Cardiff and
Toronto positions agree. We give below the positional disagreements
of all sources for which the disagreement is $>$2 arcsec and
also note sources which are particularly close to each other.
\medskip
\parindent=0pt

{\bf CUDSS 14.1:} This source is the brightest source in our survey of
the 14-hour field. It is only 0.9 arcsec away from a $\mu$Jy radio source,
with which it is clearly associated (\S 5.2). In Paper II we showed that
the optical/IR counterpart to this submillimetre/radio source is a faint
red galaxy with $\rm K_{AB} \simeq 21$.
Millimetre interferometry, which confirms the
identification proposed in Paper II, and deep optical/IR imaging and
spectroscopy
of this source are described in Paper III (Gear et
al. 2000).
\smallskip

{\bf CUDSS 14.2:} This is the second brightest of the sources in the
14-hour field. In Paper II we found a possible optical counterpart, although
the probability of this being a chance coincidence was quite high. The
new submillimetre position is further away from the optical position,
increasing the probability that the two objects are unrelated. In this
case there is no radio source to help determine the optical/IR counterpart.

\smallskip

{\bf CUDSS 14.3:} This source is 2.7 arcsec away from a $\mu$Jy radio source,
with which it is clearly associated (\S 5.2).

\smallskip

{\bf CUDSS 14.4:} This is close to CUDSS 14.13 and this
pair of sources is one of the two worst cases of
confusion in the survey, the other being the pair
14.7/14.10. However, the Cardiff and Toronto positions
for 14.4 are in good agreement.

\smallskip

{\bf CUDSS 14.5:} This is close to CUDSS 14.9, but
the two positions agree well.

\smallskip

{\bf CUDSS 14.7:} This is close to CUDSS 14.10 
and this
pair of sources is one of the two worst cases of
confusion in the survey, the other being the pair
14.4/14.13. However, the Cardiff and Toronto positions
for 14.7 are in good agreement.

\smallskip

{\bf CUDSS 14.8:} The Cardiff and Toronto positions disagreed by 2.8 arcsec
for no obvious reason.

\smallskip

{\bf CUDSS 14.9:} This is close to CUDSS 14.5, but the SCUBA position
is only one arcsec from a radio source (\S 5.2), showing that the
SCUBA position is accurate. 

\smallskip

{\bf CUDSS 14.10:} This is close to CUDSS 14.7
and this
pair of sources is one of the two worst cases of
confusion in the survey, the other being the pair
14.4/14.13. However, the Cardiff and Toronto positions
for 14.10 are in good agreement.

\smallskip
{\bf CUDSS 14.13:} This is close to CUDSS 14.4
and this
pair of sources is one of the two worst cases of
confusion in the survey, the other being the pair
14.7/14.10.
The Cardiff and Toronto positions agree well, but in this case there
is a galaxy 5.9 arcsec away from the
SCUBA position which is a radio source and ISO source (\S 5) and which
has a very disturbed optical structure (as shown by HST imaging). As 
a statistical
argument (\S 5) shows that the chance of this galaxy not being the counterpart
to the SCUBA source is low, as this is one of the two worst cases of confusion,
and as 
the investigation of the effects of confusion (\S 4.3) shows 
that offsets this large
will happen in $\simeq$10\% of cases, we conclude 
that the SCUBA position is wrong by
this amount. The redshift of the galaxy is 1.15.

\smallskip

{\bf CUDSS 14.17:} The Cardiff and Toronto positions disagree by 5.3 arcsec,
indicating that the position in Table 2 is quite inaccurate. This is probably
caused by the source being close to the edge of the field, where the noise is
varying rapidly with position. This source is 10.3 arcsec away from an ISO 15$\mu$m source
(\S 5), and although we would not consider it likely that the ISO galaxy is
associated with the SCUBA source, the large disagreement between the
two SCUBA positions makes it at least possible.

\smallskip

{\bf CUDSS 14.18:} This source is 2.0 arcsec 
away from a $\mu$Jy radio source (\S 5.2) and an ISO source (\S 5.1).
The probabilities of these being chance coincidences are very low.
The redshift of the galaxy associated with the radio/ISO/SCUBA source
is 0.66.

\smallskip

{\bf CUDSS 14.19:} The Cardiff and Toronto positions disagree by 4.5 arcsec,
indicating that the position in Table 2 is quite inaccurate. This is probably
caused by the source being close to the edge of the field, where the noise is
varying rapidly with position.

\parindent=20pt

\subsection{The Effects of Confusion and Catalog Reliability} 

We investigated the effects of confusion \cite{peter,condon} and noise
by generating Monte-Carlo simulations of our field and
then using the technique used for the real image (\S4.1)
to find the sources. 
The simulations are based on the 14$^h$ source counts. We have
assumed that, at high flux densities,
the integral source counts have the form
$N(>S) = N_0 S^{-\alpha}$, with $N_0$ and
$\alpha$ chosen to match the
14$^h$ counts; while at flux densities below a transition
flux density, $S_t$, we have assumed that the source
counts have the form $N(>S) = N_1 ln(S)$, with $N_1$ chosen
so that the source counts at low and high flux densities
are the same at $S = S_t$, and $S_t$ chosen so that the total
flux density from all the sources equals the cosmic background
radiation at 850$\mu$m \cite{fix}. As the first stage in
the Monte-Carlo experiment, we used these source counts to
generate five representations of the 14$^h$ field with no
noise component; for these fields, errors in the recovered
fluxes and positions will be entirely 
the effect of source confusion.
As the second stage, we generated five noise images, as in 
\S 4.1, and then used the assumed source counts to add sources
to these images; for these fields, errors in the recovered
fluxes and positions will be the effect of both confusion
and noise.

Figure 4 shows the results. The top row of plots in this montage
are for the fields with no noise and the bottom row
are for the fields containing noise. 
The bottom row of plots is most relevant for considering
the present reliability of the catalog, although a comparison of
the two rows shows that it is a combination of noise
and confusion which produces the effects we consider below.
The bottom lefthand plot shows
the flux of the source put in the simulation (the input
flux, $S_{in}$) plotted
against the flux recovered by the source-finding technique (output
flux, $S_{out}$). This plot shows that our sample is
surprisingly complete. A sample with $S_{out} > 3$ mJy contains
90\% of the sources with input fluxes brighter than
this. However, the less gratifying result is that the fluxes
of the sources in a sample will tend to be biased upwards;
although the effect of confusion and noise can either increase
or decrease the flux of a source, a flux-limited sample will
preferentially contain sources whose flux densities have been
boosted. 
The sizes of boost factors ($S_{out}/S_{in}$) 
can be seen most easily in the bottom righthand plot, where this
is plotted against the difference between input and
output position.
The median boost factor
is 1.44 with a large scatter about this value.

Apart from biasing the fluxes of individual sources,
this effect will produce a bias in the source counts,
which has cosmological implications (\S 6.1).
A comparison of the source counts produced from
our recovered sources with the input source counts shows
that the slope of the counts is unaffected but that the 
integral counts are shifted in flux by the average boosting
factor. 
The correct source counts to use as the
input to our simulation are the true source counts, whereas we
necessarily had to use the measured source counts. 
Since the measured source counts
will have been boosted from the true counts, our simulation will
have overestimated somehwat the effect of confusion; thus
the boosting factor we have deduced is
strictly an upper limit.

The middle plot in the bottom row shows the difference between
input and output positions plotted against output flux.
Knowledge of the 
size of the positional errors is of course crucial
for any attempt to determine the optical/IR counterparts
to the SCUBA sources.
The plot shows that 19\% of the SCUBA sources
have positional differences greater than 6 arcsec. This means
that if one searches for the optical/IR counterpart to a SCUBA
source within a circle of radius 6 arcsec centered on the
SCUBA source (we used five arcsec in Paper II), the true
counterpart 
will lie outside this circle 19\% of the time.
This is strictly an upper limit for the reason discussed above.
However, we also investigated this issue
using a separate set of
Monte-Carlo simulations, in which we placed artificial sources
on our real image, and then tried to recover the sources
using our standard algorithms. In these simulations, we found
that 12\% of the sources had positional differences greater
than 6 arcsec. The combination of these two sets of simulations
imply between 10 and 20\% of the sources have positional
differences greater than 6 arcsec.

Some insight into the nature of the sources with large
positional discrepancies comes from the two 
righthand plots. A comparison of the two shows
that there are almost three times as many sources with
positional
offsets greater than 6 arcsec when noise 
is added to the simulations as
when there is only source confusion. The lower plot
also shows 
that the median boost ($\simeq$2.3) for the sources with such large offsets
is much greater than the average boost factor
for the whole sample (1.44). 
This suggests that the sources with large offsets are produced
by the proximity of one or more faint sources to noise peaks; thus
the 
sources with large offsets and the $\simeq$14\% of the
sources
expected to be spurious from
Gaussian statistics (\S 4.1) are probably the same objects.

The results of these simulations are of course applicable to
many of the other submillimetre surveys. Our survey contains
about one source per forty beams, which is the traditional limit at which
confusion begins to becomes important \cite{peter,condon}. 
Deeper surveys, such as that of the
Hubble Deep Field \cite{dave} are beyond this limit, which may explain
why there has been so much disagreement about the
optical/IR counterparts to the SCUBA sources in this field 
\cite{dave,eric,downes}. 

\section{The Correlation With Other Surveys}

\subsection{ISO}

The 14$^h$ field has been surveyed with ISO at 6.75 $\mu$m \cite{flor1}
and at 15 $\mu$m \cite{flor2}. We looked for objects that might be both
SCUBA sources and ISO sources by looking for 15 $\mu$m sources that fall
close to a SCUBA source. 
If an ISO source lies $d$ arcsec from a SCUBA source, the probability
of it not being related to the SCUBA source is $1 - e^{-\pi n d^2}$, in which
$n$ is the surface density of ISO sources. The low surface density of ISO
sources means that it is possible to consider a much larger search radius
than is possible when looking for optical counterparts (Paper II) and
we chose a search radius of 10 arcsec. The
simulations described in the last section show that the true
positions of virtually all the SCUBA sources should lie within 10 arcsec
of the measured position.
The ISO positions, of course, have errors themselves
($\simeq$3.7 arcsec, Flores et al. 1999b),
but these are rather less than the SCUBA errors, and in the many cases where there
is an optical counterpart \cite{flor2} we have used
the position for this rather
than the ISO position. Thus the SCUBA positional errors are the dominant source
of error in this analysis, and since we are using a search radius that
is large enough to allow for this,
we should not have missed any genuine SCUBA-ISO association.

Table 4 lists the two ISO galaxies that fall within
10 arcsec of SCUBA positions and one ISO galaxy that is marginally
outside this search radius. CUDSS 14.13 and 14.18 are almost certainly
associated with ISO sources (ISO 0 and 5); the probability of a chance coincidence
is small and the two ISO sources are the brightest 15$\mu$m sources
in Flores et al.'s list (see \S 6.3 for why this is a supporting argument). 
The positional difference between CUDSS 14.13 and the
galaxy associated with ISO 0 is quite large (5.9 arcsec) but the SCUBA source is confused
with another (\S 4.2) and thus the positional accuracy may be worse than
usual. This field is quite interesting because there is a cluster
of five ISO sources apparent on the image of Flores et al., and the coincidence
of this cluster with the two confused SCUBA sources suggests that we may
have found a cluster of dusty galaxies at a redshift of 1.15, the redshift
of the galaxy associated with ISO 0. The third association is more speculative;
the probability of the association being incorrect is only 8\%, but given the
size of our sample of SCUBA 
sources we would expect to find at least one unrelated ISO source
falling this close to one of the SCUBA sources.
We have only listed this association as a possibility because the SCUBA position
is known to be inaccurate (\S 4.2), and henceforth we assume that the
association is not genuine.
 
Only two of the 19 SCUBA sources are ISO 15$\mu$m sources, and of the
$\simeq$50 15 $\mu$m sources in this field only two are submillimetre
sources.
This confirms
earlier claims that the ISO and SCUBA surveys are finding different
sets of objects \cite{dave,pap2}.

\subsection{Radio}

The 14$^h$ field has been surveyed with the VLA at 5 GHz and slightly
less deeply at 1.5 GHz \cite{fom}. We looked for associations between
SCUBA sources and radio sources detected
at $>$3$\sigma$ on the original
5GHz radio image (kindly supplied by Dr. E. Fomalont). 
Because of the higher surface density of radio sources relative to
the ISO sources (\S 5.1), it proved necessary to restrict the search radius
to six arcsec to avoid too many false associations; this then creates
the opposite problem that $\simeq$10-20\% of the genuine
associations will be missed (\S 4.3). 
The results are given in Table 5.
Of the 19 SCUBA sources in our catalog,
16 fall in the useable area of the 5GHz radio image, and of these there
are five which have radio sources within 6 arcsec. One of the
SCUBA sources, 14.13, has two radio sources at about the same distance.
We have assumed that the brighter radio source is the genuine association,
because it is coincident with the galaxy associated with ISO 0 (\S 5.1). 
For the SCUBA sources which lie outside the useable area of the
5GHz map, we obtained upper flux limits from the 1.5GHz map, which
we re-reduced.

The percentage of the SCUBA sources which are detected at radio wavelengths
is 31\% if only the sources for which we have 5GHz data are considered
and 26\% if all the sources are included.
Both percentages are
much lower than the percentage of 71\% detected at radio
wavelengths by Barger, Cowie and Richards (2000) for a different SCUBA sample.
We discuss the reasons for this difference in \S 6.2.

\section{The Immediate Implications of the Survey}

\subsection{How much of the submillimetre background has been resolved?}

The fraction of the 850$\mu$m background radiation that can be accounted
for by simply replicating the sources found in the SCUBA survey over the
sky is of great importance, because if this fraction is high (and if the
SCUBA galaxies are also representative of the stronger
submillimetre background at shorter wavelengths---\S 1), 
then they are responsible for a significant fraction of the total
energy output of the galaxy population. To estimate this fraction, however,
it is necessary to allow for the possibility of flux boosting
(\S 4.3).

Fig. 3 shows the source counts. We have assumed that the {\it differential}
counts  
have the power-law form, $dN(<S)/dS = N_0 S^{-\alpha}$, and determined
$\alpha$ using the standard Maximum Likelihood technique
\cite{jaun}, obtaining the value $3.25\pm0.7$;
the Monte-Carlo simulations (\S 4.3)
show that this is relatively unaffected
by the effects of confusion and noise.
This value 
is consistent with the values of 3.2 (95\% confidence
limits of 2.6 and 3.9) estimated by Barger et al. (1999) and
$\rm 2.8\pm0.7$ estimated
from the cluster survey \cite{blain2}.

Using this value of $\alpha$ and an estimate
of $N_0$ from the 14$^h$ counts,
and making no correction for flux boosting,
we estimate that 
27\% of the background at 850$\mu$m (Fixsen et al.
1998) is produced by individual sources brighter than 
3 mJy. After a correction of 1.44 is made in flux to allow
for flux boosting (\S 4.3), 19\% of the background is produced
by sources with true flux densities
brighter than 2 mJy (the steepness of the source counts
means that the fraction of the background produced by sources with
true flux densities brighter
than 3 mJy falls to 8\%).
The deepest blind survey is of the
Hubble Deep Field \cite{dave}, which had a sensitivity
limit of 2 mJy and which resolved about 30\%
of the background into individual sources (this is not significantly
higher than our value because the HDF counts were rather lower than
ours---Fig. 3). Flux boosting should be at least as bad in the HDF, and
if the same correction is made, 21\% of the background is produced
by sources with true flux densities
brighter than 1.4 mJy. There
is some information about the source counts fainter than 2 mJy (Fig. 3),
but this information consists of a fluctuation analysis in the HDF
and three sources in the cluster lens survey with unlensed
flux densities less than 3 mJy \cite{blain2,blain3}. Therefore,
a conservative conclusion would be that 20\% of the background
at 850$\mu$m has been resolved into individual sources.

\subsection{The radio-submillimetre relation: structures and redshift
distributions}

Radio observations of SCUBA survey fields are important for
three reasons: to provide more accurate positions; to determine
whether AGN or young stars are heating the dust; to provide
redshift estimates. The physical basis of the usefulness of
radio observations
is that at low redshift the far-infrared and radio emission
from star-forming galaxies are tightly correlated (Helou, Soifer
\& Rowan-Robinson 1986; Devereux and Eales 1989),
presumably because the dust is being heated by high-mass stars, which
then rapidly form supernovae, generating the relativistic
electrons necessary for radio emission. Not only are the dust
emission and radio emission correlated globally but they are also
correlated spatially \cite{me} and thus high-resolution radio
observations are a way of inferring whether the dust emission
is extended, as would be expected for star-forming galaxies,
or unresolved, as would be the case for AGN.
The shape of the radio-submillimetre spectral
energy distribution for a typical star-forming galaxy means that,
in the absence of evolution,
the radio to 850$\mu$m flux ratio will depend
strongly on redshift, and Carilli \& Yun (1999) have suggested
that this ratio might be
a good redshift indicator.

The most basic things the radio observations provide are
more accurate positions for individual sources and information
about the overall accuracy of the SCUBA positions. Of the five
SCUBA-radio associations, the median difference between
the radio and submillimetre positions is 2.0 arcsec (Table
5). This may be compared with the predictions of the Monte-Carlo
simulations (\S 4.3) shown in the bottom middle panel
of Fig. 4. Discarding the Monte-Carlo sources with positional errors
greater than 6 arcsec, since the radio-SCUBA associations would
have been missed completely (\S 5.2), the predicted median error is 3.5 arcsec.
Although the number of sources is small, the fact that
the median observed difference is rather smaller than the
median predicted difference suggests that
the Monte-Carlo simulations did give a conservative estimate
of the positional errors.

Of the five radio sources, four have
angular sizes of $\simeq$ 1-2 arcsec \cite{fom}. 
Sources this size would only have barely been resolved by the
VLA beam but, if correct, this result
is interesting for two reasons.
First, it
suggests that these are not AGN but star-forming galaxies. Second,
the implied physical sizes
($\rm 10h_{50}^{-1}$ kpc for $z > 1$)
are much larger than the sizes of the starburst
regions in nearby ULIRG's \cite{me}, and so
this may be the first indication that the
SCUBA sources are objects in which galaxy-wide starbursts are occurring, rather
than the nuclear starbursts seen in low-redshift ULIRG's.

Carilli \& Yun (1999) used models to investigate
how the ratio of submillimetre to radio flux 
should depend on redshift for star-forming galaxies.
Dunne, Clements and Eales (2000) have recently used
the radio, submillimetre, and far-infrared
data for the 104 galaxies in SLUGS (\S 1) to 
estimate how this ratio should depend on redshift for
real star-forming galaxies, and in particular how 
the dispersion in this ratio should depend on redshift,
since this is crucial for determing the accuracy of redshift
estimates made with this technique.
Fig. 5 shows their estimate of how this ratio should
depend on redshift, and of the
$\pm$1$\sigma$ uncertainty
in using this ratio to estimate 
redshifts. Also plotted in the figure are the submillimetre
to radio ratios for all the galaxies detected in SCUBA
surveys which have spectroscopic redshifts and useful radio
data. Of the six galaxies plotted, five are in excellent agreement
with the predictions, which is additional evidence that
the SCUBA galaxies are mostly star-forming galaxies rather than
AGN.

We have used these curves to estimate redshifts and redshift limits
for our sample.
The curves are calibrated using 1.4-GHz fluxes, and we have
converted our 5-GHz fluxes to this frequency using the
spectral indices listed in Fomalont et al. (1991).
For the SCUBA sources with only radio upper limits we have
converted the 5GHz upper limits to 1.4GHz upper limits
using a radio spectral index of 0.7, typical of the
sources found in deep 1.4GHz surveys \cite{eric2}. We have used
these upper flux limits and the median curve in Fig. 5 to estimate
lower redshift limits for these sources.
The redshift estimates and limits are given in Table 5.

We have also used Fig. 5
to estimate redshifts for two other samples: (i) the
cluster lens survey \cite{smail3}; (ii) the sample of sources
with
$\rm S_{850 \mu m} > 6$mJy 
from the Hubble Flanking Fields (HFF; Barger, Cowie and
Richards 2000).
Fig. 6 shows the redshift distributions for the three samples. 
Much of the information in all the distributions consists of
limits rather than measurements, and
so we have investigated the statistics of these distributions
using the ASURV Rev 1.2 package \cite{lavalley}, which
implements the techniques for treating censored
data described in Feigelson \& Nelson (1985). The
results are given in Table 6. 
The median redshift for the 14$^h$ sources is 2.05.
Although the large number of limits for this sample mean that
this estimate is highly uncertain (shown by the
ASURV package's failure to produce 95\% error limits 
for this sample---Table 6), it is 
close to the estimated median redshift of the optical counterparts of the 
first 12 sources in our SCUBA survey (Paper II).
The median redshifts for all the
samples lie in the range 1.5 to 2.5, with the means
lying in the range 1.9 to 2.7. We compared the distributions
for the individual samples  using the Peto-Prentice
generalised Wilcoxon test, finding that the differences
between them are not significant.
If all the samples are combined together, the median redshift
is 2.4, with 95\% confidence that it lies between 1.9 and 3.5.

Finally, we consider the reason for the different radio detection rates
for our sample and that of the HFF sources. Barger
et al. detected five out of seven SCUBA sources, giving a detection rate
of 71\% compared with 31\% for the 14$^h$ sources (\S 5.2). The 
radio observations were of similar
sensitivity, but the 850$\mu$m flux limit of Barger et al. was significantly
brighter than our limit; thus, at a given redshift, Barger et al.
would have been able to detect sources with a higher ratio of submillimetre to
radio flux. This must be part of the explanation but part of it may also
be the small numbers of sources involved. 
Although there is no significant
difference between any of the samples, 
the median
redshift of the HFF sources is the lowest of the three samples
(Table 6); and since the lower the redshift, the lower the
predicted submillimetre to radio flux ratio, the high radio
detection rate for the HFF sources may partly be a statistical
fluctuation.

\subsection{The spectral energy distributions of high-redshift dusty galaxies}

We investigated the significance of the 15$\mu$m/850$\mu$m and
450$\mu$m/850$\mu$m flux ratios of the galaxies found
in the SCUBA and ISO surveys by comparing the measured ratios
and limits with the predictions for three standard spectral
energy distributions. The first two are taken from Schmitt et al. (1997),
who list average spectral energy distributions (SED) for a number of types
of galaxy. We have used the SEDs for spirals and for high-extinction
starbursts (SBH). Since Schmitt et al. did not have access to many submillimetre
data when compiling these SEDs, we have created new SEDs at wavelengths
greater than 60$\mu$m, matching these on to the listed SEDs at this
wavelength. For the spiral galaxy, we used the two-component
fit to the far-infrared and submillimetre data for NGC 891 of
Alton et al. (1998). For the starburst, we assumed a dust temperature
of 48K and a dust-emissivity index of 1.3, a good fit to the data
for the archetypical starburst M82 (Hughes, Gear and
Robson 1994). As the SEDs in
Schmitt et al. are based on observations of galaxies that appear
in the IUE archive, even their high-extinction starbursts are
likely to have much less dust extinction than the ULIRGs revealed
by IRAS. Therefore, as a third standard SED, we used the SED of the
archetypical ULIRG, Arp 220. At wavelengths greater than 60$\mu$m,
we assumed a dust temperature of 42.2K and a dust emissivity index
of 1.2, which is a good fit to the submillimetre and far-infrared
fluxes (Dunne et al. 2000). We extended this SED to shorter wavelengths
by interpolating
between the flux densities of Arp 220 measured by IRAS and with
ground-based telescopes (Carico et al. 1988). 

Fig. 7 shows the 15$\mu$m/850$\mu$m flux ratios 
and limits for
the ISO and SCUBA galaxies plotted against their redshift,
together with the predictions of the three SEDs. The ISO galaxies
are the ones with spectroscopic redshifts in Catalogue 1 of
Flores et al. (1999b), and for simplicity the lower limits on the flux
ratio
have been calculated assuming an upper limit to the 850$\mu$m
flux density of 4 mJy. The upper limits to the flux ratio for the
SCUBA galaxies have similarly been calculated using an upper
limit to the 15$\mu$m flux of 0.18 mJy. Only two of the
SCUBA galaxies have spectroscopic redshifts, and for the remainder
we have used the redshift estimates or limits obtained from the
radio method (\S 6.2). 

There are a number of straightforward conclusions that one can reach
from this diagram. The flux ratios of the ISO
galaxies are consistent with the predictions
of the spiral and SBH SEDs 
but are not consistent in most cases with the prediction of the Arp 220 SED.
The explanation for this is quite simple. Arp 220 has a very steep
SED
between the near-infrared and 60$\mu$m, presumably because the
extinction in this object is so high that even the long-wavelength
emission is significantly absorbed. Therefore, its 15$\mu$m flux
falls rapidly with redshift. Thus the ISO mid-infrared surveys are
picking up galaxies in which the extinction is much more modest than
in the ULIRGs discovered with IRAS. Note, this shows the
care needed in modelling the results of the mid-infrared
surveys; it is not simply a matter
of extrapolating models based on IRAS results (see Roche and Eales 1999
for further discussion). 
A consideration of the curves
also shows why it is the two ISO galaxies with the highest
15$\mu$m flux densities which are associated with SCUBA sources
(\S 5.1). Given the high flux ratios predicted by the spiral
and SBH models at $\rm z < 1$, only a galaxy with a very high
15$\mu$m flux would also be detected at 850$\mu$m.

In contrast to the ISO galaxies, the limits for the SCUBA sources are
consistent with the Arp 220 SED, are always inconsistent with
the spiral SED, and generally disfavour the SBH SED. These
results show that the mid-infrared and submillimetre surveys
are picking up different classes of galaxy. 

Fig. 8 shows the ratio of 450$\mu$m to 850$\mu$m flux for
the sources in this paper and those in Paper I and II which
have redshift measurements, estimates, or limits.
Apart from the predicted curves based on the SEDs discussed
above,
the plot also shows a predicted curve based
on the SED of the high-redshift galaxy IRAS 10214$+$4724,
which has an estimated dust temperature of 80K and
a dust emissivity index of 2 \cite{downes2}.
The first conclusion one can draw 
is
that whereas the SEDs of the SCUBA galaxies are 
always consistent with the first three models, they are sometimes 
inconsistent
with an SED of this kind (a single temperature
and a dust emissivity index of 2), 
something which is also true of the galaxies in SLUGS (Dunne et al. 2000).
The predictions for all the SEDs initially remain constant
with redshift, because at low redshift
both the 450$\mu$m and 850$\mu$m
observations are sampling the Rayleigh-Jeans tail
of the SEDs; it is only when, in the rest-frame of the galaxy,
the 450$\mu$m observations are getting close to the peak
of the SED that the predictions begin to fall.
This occurs at a lower redshift for a lower dust temperature, which
explains why the prediction for the spiral SED, 
which includes emission from 15K dust,
starts to fall at a lower redshift than the other two.
The 3$\sigma$ limit on the average 450$\mu$m/850$\mu$m
flux ratio for the SCUBA galaxies (\S 4.1) is shown by the
horizontal line. If these galaxies are typically at redshifts
of $\sim 2$, as suggested by the radio method (\S 6.2), then this
limit implies the dust in the SCUBA galaxies is colder
than that found in Arp 220 and M82.
The alternative is that the galaxies are generally
at much higher redshifts than indicated by the radio method. 
The first possibility seems most likely because there is no
reason to doubt the radio method and is of great
interest because
it is the second piece of evidence (the radio sizes being the other)
that SCUBA galaxies are not simply high-redshift ULIRGs.

\section{The Cosmological Significance of the SCUBA Sources}

This section describes modelling of the SCUBA population
with the aim of answering some of the key questions
about this population (\S 1):
(1) Are the high-redshift
galaxies discovered at optical wavelengths through the Lyman break
technique a completely separate population from the SCUBA galaxies?
(2) Did the universe
have a Dark Age in which 10 times as much star formation was hidden
by dust as appears at optical wavelengths?
(3)
Does the submillimetre luminosity density keep on increasing with
redshift or does it decrease above $\rm z \sim 2$? 
(4) If the SCUBA
galaxies are proto-ellipticals, when exactly did ellipticals form? 
About a fifth question, whether the dust is heated by AGN or young
stars, we can say little
beyond noting
that both the radio structures
and the radio-to-submillimetre flux ratios (\S 6.2),
as well as the first deep X-ray surveys (Fabian et al.
2000; Hornschemeier et al. 2000), suggest that
it is the latter; and this is
what will be assumed through the rest of this section.
The focus of the modelling described in \S 7.2
will be on determing why there
are conflicting answers to many of these questions.
First, however, we will consider separately the first question.

\subsection{Are the Lyman-Break and SCUBA Galaxies the same Population?}

As Lyman-break galaxies are found from their properties in
the rest-frame ultraviolet and SCUBA galaxies are found from their
dust emission, one might expect that the statistical properties
of the two populations should be very different. Adelberger
and Steidel (2000), however, have recently suggested that the two
populations are essentially the same,
and that there is no evidence that a significant fraction 
of the star-formation
in the universe occurred in galaxies so heavily obscured that they
could not be detected in UV-selected surveys.
One test of this idea is to determine directly 
the submillimetre properties
of the Lyman-break galaxies.
Although  SCUBA
observations of individual Lyman-break
galaxies have resulted in only a single detection \cite{chap},
Peacock et al. (2000) have claimed a statistical detection
of the population by summing the submillimetre emission at the
positions of Lyman-break galaxies in the deep SCUBA image
of the Hubble Deep Field. 
The greatest uncertainty in this claim
is the fact that Lyman-break galaxies are highly-biased
and thus highly-clustered systems (Steidel et al.
1998; see Fig. 4 of Peacock
et al.) and if SCUBA galaxies are
similarly highly-biased highly-clustered systems following
the same large-scale structure as the Lyman-break galaxies,
the small number of SCUBA beams in the Hubble Deep Field
means that the technique of Peacock et al.
might yield a detection even if the Lyman-break galaxies
themselves contain no dust.
Nevertheless, in this section we assume the claim is correct
and investigate its
implications by predicting the
contribution of Lyman-break galaxies to the
submillimetre background and source counts.

Adelbeger and Steidel predicted the global submillimetre properties
of the Lyman-break galaxies by using correlations between bolometric 
luminosity and fluxes at UV, mid-infrared and submillimetre wavelengths
established for local starbursts. Here we adopt the simpler
approach of using the average ratio of submillimetre to optical
flux measured for
the Lyman-break population by Peacock et al. (2000).
We predicted the 850$\mu$m source counts and the
submillimetre background from the  
Lyman-break population in the redshift range
$1 < z < 5$. We used the average ratio of
850$\mu$m to optical flux for the 20 Lyman-break
galaxies in Table 1 of Peacock et al. 
and the
I-band luminosity function
given by Steidel et al. (1999) for
Lyman-break galaxies at $\rm z \sim 3$; we
assumed that there
is no cosmic evolution in the luminosity function
over this redshift range.
We have also implicitly assumed that the submillimetre
to optical flux ratio found by Peacock et al. for
the most luminous Lyman-break systems applies to the
whole population.
Even if the Lyman-break population has been detected
in the submillimetre, there is no information about their
dust temperatures; so we tried two dust temperatures,
20K and 40K, and assumed a value for the dust emissivity of
1.2, similar to that seen for the low-redshift galaxies
M82 and Arp 220 and consistent with the limited amount of
multi-wavelength submillimetre photometry for SCUBA galaxies
(\S 6.3). Fig. 9 shows the predicted submillimetre background 
and source counts. 

The predicted contribution of the Lyman-break galaxies to
the background depends critically on dust temperarure: if the
dust temperature is 40K, the Lyman-break galaxies produce most
of the submillimetre background; if the dust temperature is
20K, their contribution to the overall background
is not so important, although they do produce the entire
background at long wavelengths. 
The predicted source counts do not match the observed source counts
at high flux densities although the discrepancy is only a factor
of 5.3 in flux density (or 3.8 if flux-boosting is taken into
account). 

Thus this simple model shows that, if Peacock et al.
are correct, the Lyman-break population
might be an important component of the submillimere
source counts and background. However, there are two
important points to consider. First,
if the submillimetre to optical
flux ratio derived for the Lyman-break galaxies
applied only to the most luminous systems, then
since the background is dominated by low-luminosity
systems, this conclusion would be incorrect.
Second, 
at $\rm z < 3$, the 850$\mu$m emission is approximately
proportional
to $M_d T_d$, in which $M_d$ is dust mass and $T_d$ is dust
temperature. Thus, dust mass is as important a predictor of
850$\mu$m flux density as dust temperature; and if the
Lyman-break galaxies were massive galaxies, it would not be
surprising for the most luminous Lyman-break galaxies
to be detected by SCUBA at a flux level a few times less than 
the sources revealed in the blind surveys. 
There is at present little evidence as to the masses of the
Lyman-break galaxies, although measurements of nebular lines
in the infrared for a handful of objects
suggest that, in general, Lyman-break galaxies are not very massive galaxies
\cite{max}.

Finally, we note (as we did in \S 1) that
the optical properties of the SCUBA sources do not
in general support this hypothesis. If it
were correct, the optical/IR counterparts to the
SCUBA sources should be bright enough to be detected
in the Lyman-break surveys; and this should be especially true
of the counterparts to the brighter SCUBA sources. Consider,
however, the brightest 14$^h$ source, CUDSS 14.1 \cite{walt}. We
have estimated, using the radio technique (\S 6.2), that its
redshift is $\simeq$2. From its measured magnitudes \cite{walt}, we
estimate that if this source were moved out to a redshift of
3, 
the redshift
of the Lyman-break galaxies, its magnitude would be $\rm I_{AB} \simeq
27$. This is two magnitudes fainter than the limit of the Lyman-break surveys.

There are a large number of uncertainties in the discussion above. However,
we conclude, mainly because of the argument in the last
paragraph, that the SCUBA galaxies and the Lyman-break galaxies are
largely separate populations. We suspect that the submillimetre
detections that are beginning to be made of the Lyman-break galaxies
are largely caused by the sensitivity of the 850$\mu$m emission to dust mass;
if some of the Lyman-break galaxies are massive galaxies, one would
expect detections at roughly the level found.

\subsection{The Hidden Star-Formation History of the Universe}

We addressed the remaining questions 
by investigating how the submillimetre properties
of galaxies must change with redshift in order to reproduce the
observations: the submillimetre background, the source counts, and
the redshift distribution. The simplest types of cosmic evolution
are luminosity evolution, in which
the luminosities
of individual galaxies change with time but the total number
of galaxies remain the same, or density evolution, in which
the luminosities stay the same but the total number changes.
In the submillimetre waveband, however, density evolution
does not work, because the strength of evolution needed to reproduce
the 850$\mu$m source counts leads to much too high a submillimetre
background (Paper I; Blain et al. 1999a). We 
therefore assumed that the bolometric
luminosities of the dust in all galaxies vary with redshift in the
following way:
\medskip
$$
L_{bol} = L_0 (1+z)^p\ {\rm at}\ z < z_t, \eqno(1a)
$$
\smallskip
and
\smallskip
$$
L_{bol} = L_0 (1+z)^q\ {\rm at}\ z > z_t, \eqno(1b)
$$
\medskip
in which $z_t$ is a transition redshift. Our method was to try thousands
of combinations of $p$, $q$, $z_t$ in different cosmologies in order
to find ranges of these values consistent with the observations.
Since there are many combinations of these parameters that do give
acceptable fits to the data, there is no need to try a more complicated
model---and this method does give valuable insight into why
there is so much disagreement about the cosmological significance of the
SCUBA galaxies.

The local submillimetre luminosity function which formed
the basis of the models
was the 850$\mu$m luminosity function derived, as part of the
SCUBA Local Universe and Galaxy Survey (Dunne et al. 2000), from
SCUBA observations of 104 galaxies from the IRAS Bright
Galaxy Survey (BGS; Soifer et al. 1989). In making predictions of the
850$\mu$m source counts and redshift distribution,
rather than working from the local luminosity
function, we started from the derived spectral energy distributions
of the 104 galaxies and used standard accessible volume technques
\cite{avni}. To give a specific example, the number of galaxies
with $S_{850 \mu m} > 4$ mJy in an area of sky $A_{SCUBA}$ is
given by
\bigskip
$$
N(> 4 mJy) = \sum_i {A_{SCUBA} \int_0^{z(P_i,S_{850 \mu m} = 4 mJy)} dV
\over A_{BGS} \int_0^{z(P_i, S_{60 \mu m} = 5.24Jy)} dV }
$$
\bigskip
in which $A_{BGS}$ is the solid angle of the original sample
from which the galaxy was drawn, the sum is 
over all 104 galaxies,
and $V$ is comoving volume.
The upper redshift limits are the redshifts to which the i'th galaxy would
have been seen in the BGS and in a sample with $S_{850 \mu m} > 4 mJy$.
These were calculated using the
single-temperature spectral energy distribution (SED) for each galaxy
found by Dunne et al. to be a
good fit to the far-infrared and submillimetre
flux densities (this does
not imply that there is dust of only one temperature in
the galaxy, merely that this is an adequate representation
of the empirical SED). In the first set of models,
to implement the cosmic evolution, we assumed that 
the amplitude of the
SED of each
galaxy, but not its spectral shape, evolves with redshift
according to equation (1). 

We used a slightly more complicated approach to predict the
submillimetre background because there is a significant 
contribution to the background from
galaxies with lower luminosities than are
properly sampled by the 850$\mu$m local luminosity function
of Dunne et al.
Therefore, we used
the 60$\mu$m local luminosity function \cite{saunders} as our basic local
luminosity function, using the submillimetre data from
SLUGS to extrapolate this into the submillimetre waveband.
The dust temperature of the SLUGS galaxies
is
correlated with 60$\mu$m luminosity, with the form
\medskip
$$
T_d = 4.61 \times log_{10} L_{60} - 72.7
$$
\medskip
\noindent in which $L_{60}$ is monochromatic luminosity at
60$\mu$m in Watts Hz$^{-1}$ sr$^{-1}$. We used this
equation
and the median dust emissivity index for the galaxies
in the SLUGS (1.2, Dunne et al. 2000) 
to derive a far-infrared to submillimetre SED at each
60$\mu$m luminosity. As for the source counts, we implemented
cosmic evolution by assuming that the SED did not change its
shape with redshift, merely its amplitude.

We tried to reproduce only three sets of observations:
the overall submillimetre background, the 850$\mu$m source
counts, and the redshift distribution of the SCUBA galaxies.
In a subsequent paper we will extend the models to consider the
ISO results and also to make predictions for SIRTF and
SOFIA (Eales and Dunne, in preparation).
In attempting to reproduce the
850$\mu$m counts, we corrected the observed counts for
the effect of flux-boosting (\S 4.3); thus the observed counts at
4 mJy are assumed to represent the real counts at 2.8 mJy. 
A model was judged to provide an acceptable fit to the
observations if it satisfied the following criteria:
(i) the predicted submillimetre background was at no
wavelength more than a factor of two different from
the observed background; (ii) the predicted
850$\mu$m source
counts at 2.8 mJy were no more than 2$\sigma$
different from the observed counts at 4 mJy;
(iii) the fraction of galaxies
predicted to be at redshifts less than two in a sample
with $\rm S_{850 \mu m} (observed) > 4.0 mJy$ lay between 0.2 and 0.8.
These criteria of acceptability are quite generous and
recognise that much of the uncertainty in the submillimetre
waveband is systematic rather than statistical; the
main uncertainty in the background, for example, is whether
the subtracted forgrounds have been adequately modelled rather
than signal-to-noise. The redshift criterion is 
our admittedly subjective estimate of how uncertain this fraction is
(\S 6.2).

Fig. 10 shows the predictions of the models for
a universe with $\rm \Omega_0 = 1$ and no galaxies or equivalently
no dust beyond
a redshift of 5.
Each plot, which is for
a separate value of $z_t$, shows the low-redshift exponent
for the evolution model, $p$, plotted against the
high-redshift exponent, $q$. The greyscale shows the
fraction of galaxies predicted to lie at $\rm z < 2$ in our
canonical SCUBA sample with $\rm S_{850\mu m} (observed) > 4.0$ mJy.
The continuous line encloses models which give acceptable 
predictions for all the data; the dashed line shows the
additional region in which the models match the source
counts and the background but do not predict the right fraction
of sources at $\rm z < 2$.
For $z_t = 1, 1.5, 2$ there are many acceptable
models, with strong evolution below the transition redshift
and with either negative, or, at the most, mild positive evolution
above this redshift. For $z_t = 2.5$, the only models which fit the
data are ones with strong negative evolution
beyond the transition redshift.

We also used the models to make
predictions for a universe with $\rm \Omega_0 = 0.2$,
$\Lambda / (3 H_0^2) = 0.8$. There are no models that fit both the
850$\mu$m source counts and submillimetre background for this cosmology,
for the following reason.
The SCUBA sources are high-luminosity sources whereas there is
a significant contribution to the background from low-luminosity
sources. In this cosmology, 
models which generate enough high-luminosity
sources to reproduce the source counts, generate so many low-luminosity
sources that the predicted background is too high. We could produce acceptable
models by only including evolution for sources above some critical luminosity,
but this is beyond the scope of this paper.

In all of these models we assumed that the spectral shape
of the SED of each SLUGS galaxy is independent of redshift and
implemented the required bolometric luminosity evolution by
having the normalization of this SED change with redshift.   
It is possible, however, that the temperature of the dust in
galaxies also changes with redshift. 
In accord with our approach of exploring the full range of models
that might give acceptable agreement with the observations, we tried
the following model. 
In this variation on the basic model,
the SEDs of the SLUGS galaxies
remain constant until the transition redshift, where the temperature
of the dust is then increased by a factor of 1.5, with the normalization
of the SED scaled appropriately so that there is not a sudden jump
in bolometric luminosity. 
This abrupt change in the dust temperature is undoubtedly unphysical
but is the simplest way of investigating the possibility that the average
dust temperature at high redshift is much greater than it is today.
In the model the typical
dust temperatures of the SLUGS galaxies above the transition redshift are
45-65K, not inconsistent with the limited
multi-wavelength submillimetre photometry of SCUBA sources (\S 6.3) and
much lower than the dust temperature of 80K measured for
the high-redshift galaxy 
IRAS 10214$+$4724 \cite{downes2}.
Figs 11 and 12 show the results for this model for maximum redshifts
of 5 and 10. Both figures show that there are now acceptable models
in which the bolometric luminosity keeps on rising with redshift
above the transition redshift. The reason for this can be seen by considering
the SED of a galaxy at $\rm z \sim 3$. Increasing the dust temperature of
such a galaxy increases its bolometric luminosity by a large factor but has
a much smaller effect on its 850$\mu$m flux density which, even at
this redshift, is emitted from the galaxy on the Rayleigh-Jeans tail
of the SED. Thus the 850$\mu$m source counts do not provide a strong
constraint on this model. The increase in the bolometric luminosity
has, of course, a large effect on the predicted background; but the
increase in temperature means that the increase in the background occurs
at wavelengths of $\sim$200 $\mu$m, where the observed background is
at a maximum.

Now consider the questions raised at the start of this section,
starting with whether the submillimetre luminosity density has a
maxiumum. Figs 10-12 show
that there are acceptable models in which the
submillimetre luminosity density reaches a maximum and also ones
in which it increases monotonically with redshift.
Thus the observations at present are insufficient to decide between
these two possibilities.

Now consider the fraction of the star formation that 
is hidden by dust
at low and high redshift. 
Fig. 13 shows bolometric dust luminosity plotted against redshift
for all acceptable models.
We will assume that this represents
the energy emitted by young stars that is
absorbed by dust. 
It also necessary, of course,
to estimate the
optical/UV luminosity density associated with the young stars
that is not absorbed by dust. 
The main uncertainty in doing this is that a large part of the
optical light from galaxies
is not from young stars but from evolved stars. 
We made the simple assumption
that all the light from galaxies at wavelengths less than 5000\AA\ is
from young stars. We then estimated the optical/UV luminosity density
in the local universe using the galaxy luminosity functions for
galaxies of different morphological types
given by Folkes et al. (1999) and the
SEDs for galaxies of different morphological types given by
Coleman, Wu and Weedman (1980). At high redshift, we started with the
luminosity function given by Steidel et al. (1999) for Lyman break
galaxies at $\rm z \sim 3$. This luminosity function is at a rest-frame
wavelength of $\sim$2000\AA, and it is necessary to make some
assumption about the typical SED of a Lyman-break galaxy in the
wavelength range $\rm 2000\AA < \lambda < 5000\AA$, about which little
is known. In practice, we made two estimates of the luminosity density,
one assuming that the flux density per Hertz of a Lyman break galaxy
is flat in this wavelength range and one using the SED given for
west MMD11 by Adelberger and Steidel (2000). These estimates differed
by 30\% and we took the average of the two.

These estimates are shown in Fig. 13. At low redshift the percentage
of the light from young stars that is hidden by dust is about 50\%.
At high redshifts, there are models in which this percentage is less
but also models in which 90\% of the light from young stars is absorbed
by dust. Thus conclusions that the universe had a `Dark Age' in which
90\% of the light is absorbed by dust (Hughes et al. 1998; Barger, Cowie
\& Richards 2000) are premature, but neither can one rule out this
possibility.

Finally there is the question of the origin of ellipticals.
The colours and spectra of nearby ellipticals have traditionally
been used to argue that most of the stars in ellipticals form
at a high redshift
(e.g. Bower, Lucey
\& Ellis 1992). If most of the
stars in ellipticals formed at, for example, $\rm z > 3$, 
then
given the
quantity of metals associated with nearby ellipticals \cite{mike}, 
about half the stars that have ever formed should have formed
beyond this redshift. For each model
we have calculated the fraction of stars
that have formed by a redshift $z$ using the following formula: 
\bigskip
$$
f = { \int^{t(z)}_{t(z_{max})} L_{bol}(t') dt' \over \int^{t=0}_{t(z_{max})}
L_{bol}(t') dt'}.
$$
\bigskip
\noindent This implicitly makes the assumption that the star-formation rate
is proportional to the bolometric luminosity. Fig. 14 shows $f$
for all acceptable models.
If dust
temperature does not change with redshift, then the largest percentage
of star formation that has occurred by a redshift of three is $\simeq$15\%,
similar to the 
conclusion of  Paper II from
the redshifts of the 
optical counterparts.
If dust temperature does evolve, this percentage rises to 32\%; and the
percentage would also rise in the analysis of Paper II
if the dust temperature
were allowed to evolve in this way.
Thus we can not rule out the possibility that a significant fraction
of stars have formed by this redshift.

It is clear that there are two types of observation which would
allow substantial progress to be made. First, measurements
of the temperature of the dust in the SCUBA galaxies would
make it possible to estimate directly the bolometric
luminosity of individual sources and of the population as a
whole, which are crucial for all the questions. This should be
possible through short-wavelength observations with SCUBA and, in
the future, through observations with SOFIA and SIRTF.
Second, the predicted redshift distributions shown
in Figs 10-12 show that determining redshifts of more
SCUBA galaxies would immediately rule out many models. It 
will be possible to make progress here through deep radio observations
and millimetre/submillimetre interferometry, to obtain
more accurate positions, and through optical/IR imaging
and spectroscopy on 8-10m telescopes.

\section{Conclusions}

We have used the SCUBA submillimetre camera to survey
an area of $\rm \simeq 50\ arcmin^2$, detecting 19 sources
down to a 3$\sigma$
sensitivity limit of between 3 and 4 mJy at 850$\mu$m and
obtaining the following results:

\begin{enumerate}

\item Of the 19 sources, 16 fall in a region with
radio observations with $\mu$Jy sensitivity, and
of these five are detected at radio wavelengths.
The radio/submillmetre flux
ratios and radio sizes suggest that the dust in these
galaxies is being heated by young stars rather than AGN, although
the radio sizes also suggest that the stars are being formed
over a much larger region than is seen for low-redshift
Ultra-Luminous Infrared Galaxies (ULIRGs).

\item We have used the radio to submillimetre flux ratio
to estimate that the median redshift of the SCUBA
sources is $\sim$2. The redshift distribution is
consistent with that determined in other surveys.

\item By coadding the 450$\mu$m emission at the
positions of the 850$\mu$m sources, we obtained
a 3$\sigma$ upper limit to the average
450$\mu$m/850$\mu$m flux ratio of 1.9,
which implies that either the dust in the SCUBA
galaxies
is generally colder than in ULIRGs or that they
are generally at $\rm z >> 2$.

\item A comparison of the SCUBA image with the 15$\mu$m
image of this field obtained with ISO shows that only
two of the 19 SCUBA sources were detected at 15$\mu$m,
and, conversely, only two of the 50 ISO sources were detected
at
850$\mu$m. 
A comparison of the 15$\mu$m/850$\mu$m flux ratios
with predictions based on model spectral energy distributions
shows that this result is not simply caused by the submillimetre
and mid-infrared surveys being sensitive to different redshift ranges,
but that
the ISO surveys
will tend to have missed ULIRGs.

\item Monte-Carlo simulations of the field
show that the fluxes of sources in all SCUBA surveys will
have been significantly
biased upwards by the effects
of source confusion and noise, leading to an overestimate
of the fraction of the 850$\mu$m background that has been
resolved by SCUBA. We reach the
conservative conclusion that $\simeq$20\% of the background
at 850$\mu$m has been resolved by SCUBA.
The simulations have also been used
to quantify the effects of confusion on
source positions.

\end{enumerate}

We have used the SCUBA Local Universe and Galaxy Survey and
simple evolution models to address the major questions
about the SCUBA sources: (1) what fraction of the star formation
at high redshift is hidden by dust? (2) Does the submillimetre
luminosity density reach a maximum at some redshift? (3) If the
SCUBA sources are proto-ellipticals, when exactly did ellipticals
form? The observations are not yet good
enough, however, to answer these questions: there are acceptable
models in which, at high redshift,
10 times as much star formation is hidden by
dust as is seen at optical wavelengths, but also 
ones in which the star formation hidden by dust is less than
that seen optically; there are acceptable models in which the
submillimetre luminosity density reaches a peak, but also
ones in which it continues to rise with redshift; finally there
are acceptable models in which very little star formation occurred
before a redshift of three (as might be expected in models
of hierarchical galaxy formation), but also ones in which
30\% of the stars have formed by this redshift.
The models show that the keys to answering these
questions are 
measurements of the
dust temperatures and redshifts of the SCUBA sources.

\acknowledgments

We are grateful to the many members of the staff of the
Joint Astronomy Centre, in particular Wayne
Holland, that have helped us with this
project. We thank Dr. E. Fomalont for allowing us to use
his deep VLA image.
Research by Simon Lilly is supported by the Natural
Sciences and Engineering Research Council of Canada and by the
Canadian Institute of Advanced Research.
Research by David Clements, Loretta Dunne, Stephen Eales,
and Walter Gear is supported by the Particle Physics and Astronomy
Research Council. The JCMT is operated
by the Joint Astronomy
Center
on behalf of the UK Particle Physics and
Astronomy Research Council, the Netherlands Organization for Scientific
Research and the Canadian National Research Council.



\appendix





\clearpage



\figcaption{Bolometric luminosity verses redshift for a source
with an 850$\mu$m flux density of 4 mJy and a dust temperature
of 20, 30, 40, 50 and 60K. The dashed lines are for a universe with
$\rm \Omega_0 = 0$ and the continuous lines for $\rm \Omega_0 = 1$.
The thick dashed line shows
the bolometric luminosity of the archetypical ULIRG, Arp 220.}

\figcaption{Images of the 14$^h$ field. Each
image is approximately 6.9 times 6.4 arcmin$^2$. An edge
region containing artefacts, a well-known phenomenom
of the SCUBA map-making process, has been removed. The raw image convolved
with a Gaussian of size (FWHM) 10 arcsec, which degrades
the resolution slightly (from 14 to 18 arcsec) but
reveals the faint sources more clearly, is shown in (a).
The image after all the 19 sources have been removed is
shown in (b). An artificial image showing the 19 sources
at the 14-arcsec resolution of the telescope is shown
in (c). The noise image described in \S 4.1 is shown
in (d). On this image, 
the two dashed contours indicate noise levels of 0.8 and 0.9 mJy;
the continuous lines indicate noise levels of 1.1, 1.2, 1.3...mJy.}

\figcaption{Integral source counts at 850$\mu$m. The key to the symbols
is shown in the figure.
The error bars in each case are Poisson error
bars and, as the counts are integral counts,
the error bars for a particular survey are not
independent.}

\figcaption{Results of the Monte-Carlo simulations of
the 14$^h$ field described
in \S 4.3. The top row of plots is for the simulations
in which the artificial images only contain sources and
the bottom row is for the images which also contain noise.
The lefthand plot in both rows shows the 
true flux of the source (input flux) versus the flux measured by
the source-detection algorithm (output flux). 
The sloping line shows where the input and output fluxes
are the same and the dashed lines indicate notional catalog
limits of 3 mJy. 
The middle plot in each row shows the difference in arcsec
between the
true position of a source and the position determined by
the source-detection algorithm plotted against output
flux. The righthand plot shows this positional difference
plotted against the ratio of the output flux to the input flux
(the `flux boost') For clarity, in the bottom row only
sources with either input or output fluxes brighter
than 3 mJy have been included.}

\figcaption{The 
predicted ratio of 850$\mu$m to 1.4 GHz flux
for star-forming galaxies from Dunne, Clements and Eales (2000).
The thick line shows the average value
of this ratio 
and the thin lines
show the $\pm$1$\sigma$ dispersion.
The measured ratios are shown for all SCUBA galaxies
with spectroscopic redshifts and radio detections
(this paper; Barger, Cowie \& Richards 2000;
Smail et al. 2000).}

\figcaption{Redshift histograms for 
(a) the sample of Smail et al. (2000), (b) the
sample of Barger, Cowie
and Richards (2000) with $\rm S_{850 \mu m} > 6mJy$,
(c) the sources in the 14$^h$ field.
The hatched parts of the histograms show the
redshift limits
obtained using the
radio method described in \S 6.2. 
The redshift ``measurements'' are either 
estimates from the
radio technique or in a few cases are spectroscopic redshifts.}

\figcaption{The ratio of 15$\mu$m flux to 850$\mu$m flux
for the galaxies in
the 14$^h$ field detected
in either this survey or in the 15$\mu$m survey
of Flores et al. (1999b). The two measurements
are for the two galaxies detected in both surveys.
The lower limits are for the galaxies with spectroscopic
redshifts in Catalogue 1 of Flores et al. The upper
limits are for the sources detected at 850$\mu$m which
were not detected at 15$\mu$m; none of these has a spectroscopic
redshift, so the redshift or redshift limit 
estimated from the radio method has
been used (\S 6.2). The lines show the predictions of the models
described in the text (\S 6.3): the continuous line is
the prediction of the high-extinction starburst (SBH) model;
the dot-dash line is for the spiral galaxy model; and the dashed
line is the prediction of the model based on the spectral
energy distribution of Arp 220.}

\figcaption{The ratio of 450$\mu$m flux to 850$\mu$m flux
redshift for the galaxies detected
in either the 14$^h$ field (this paper) or in the
$3^h$ and 10$^h$ fields (Paper I). Only objects
for which there is some redshift information
(a spectroscopic redshift or an estimated redshift
or limit) have been plotted: the filled symbols
show galaxies which have spectroscopic redshifts;
the squares show galaxies which have estimated redshifts,
either from the radio method (\S 6.2) or from optical/IR
colours (Paper II); the circles show galaxies for which
there are redshift limits estimated using the radio technique.
The horizontal line shows the 3$\sigma$ upper limit on
the average 450$\mu$m to 850$\mu$m flux ratio for the
sources detected in the 850$\mu$m survey, derived by
summing the 450$\mu$m emission at the position of
each 850$\mu$m source (\S 4.1). The lines show the predictions of the models
described in the text (\S 6.3): the continuous line is
the prediction of the high-extinction starburst (SBH) model;
the dot-dash line is for the spiral galaxy model; the dashed
line is the prediction of the model based on the spectral
energy distribution of Arp 220; and the dotted line is
the prediction of the model based on the
spectral energy distribution of IRAS 10214$+$4724.}

\figcaption{Submillimetre properties 
predicted for the Lyman-break galaxies
by the model described in the
text (\S 7.1), which is based on the submillimetre
detection of this population by Peacock et al. (2000).
The upper plot shows the predicted
submillimetre background
from this population on the assumption of a dust
temperature of 20K (continuous line) or 40K (dashed
line). The thick dot-dash line shows the measured
submillimetre background (Fixsen et al. 1998). 
The bottom plot shows the predicted 850$\mu$m source
counts (continuous line; the predicted counts do
not depend on dust temperature). The symbols show
the measured counts and the key is the same as in Figure 3.}
 
\figcaption{Predictions of the evolution models described in the
text (\S 7.2). In these models the bolometric luminosity of
the dust in a galaxy
is assumed to vary as $(1+z)^p$ at $z < z_t$ and as $(1+z)^q$
at $z > z_t$. Each plot is for a different value of $z_t$ and
shows $p$ plotted against $q$. The
continuous line in each plot encloses the models which give an acceptable
fit to the 850$\mu$m source counts and the submillimetre background
(`acceptable' is defined in the text)
and also predict that the fraction of sources 
with $\rm S_{850 \mu m} > 4 mJy$ that have
$\rm z < 2$ falls between 20 and 80\%, our admittedly subjective
estimate of how uncertain this quantity is.
The dashed line shows the additional region of acceptability if
this redshift criterion is ignored.
The greyscale further shows the percentage of sources predicted to
lie at $\rm z < 2$; 
the ten levels of the greyscale
are:
0 to 10\%, 10-20\%, 20-30\%, 30-40\%, 40-50\%, 50-60\%,
60-70\%, 70-80\%, 80-90\%, 90-100\%. 
This montage is for
a maximum redshift of 5.0 and a density constant ($\rm \Omega_0$) of
1.}

\figcaption{The same as Figure 10 except that the dust temperatures
of all galaxies are assumed to increase by a factor of 1.5 at $\rm z=z_t$.
Note that there are no acceptable models for $\rm z_t = 1$.}

\figcaption{The same as Figure 10 except that the dust temperatures
of all galaxies are assumed to increase by a factor of 1.5 at $\rm z=z_t$
and the maximum redshift has been raised to 10.}

\figcaption{Bolometric luminosity density in the universe
versus redshift
for all models which are in agreement with the observations.
The horizontal lines show the optical/UV luminosity
density in the universe at $\rm z = 0$ and $\rm z \sim 3$
estimated using the methods described in the text.}

\figcaption{The fraction of stars that have formed by a redshift
$z$ for all the models which are in agreement with the observations.}

\clearpage





\clearpage

\begin{deluxetable}{rrrr}
\footnotesize
\tablecaption{Observing Log \label{tbl-1}}
\tablewidth{0pt}
\tablehead{
\colhead{Date} & \colhead{Integration Time}   & \colhead{$\tau_{850 \mu m}$}   &
\colhead{$\tau_{450 \mu m}$}
}
\startdata

1998 March 5 &  9600s & 0.21-0.25 & 1.11-1.39 \\
1998 March 6 & 12160s & 0.13-0.14 & 0.57-0.68 \\ 
1998 March 7 & 12160s & 0.17-0.23 & 0.84-1.23 \\ 
1998 March 8 &  9600s & 0.20-0.22 & 1.03-1.16 \\
1998 March 9 &  9600s & 0.20-0.22 & 1.09-1.20 \\
1998 March 10 & 12800s & 0.16-0.17 & 0.72-0.77 \\
1998 March 11 & 9600s & 0.17-0.17 & 0.83-0.85 \\
1998 March 12 & 9600s & 0.18-0.19 & 0.89-0.90 \\
1998 March 13 & 12800s & 0.20-0.22 & 1.19-1.23 \\ 
1998 March 15 & 9600s & 0.22-0.29  & 1.20-1.65 \\ 
1998 June 16 & 11392s & 0.20-0.25 & 1.24-1.53 \\
1998 June 17 & 9600s & 0.17-0.18 & 0.88-0.95 \\
1998 June 18 & 12800s & 0.23-0.29 & 1.21-1.73 \\
1998 June 19 & 8320s & 0.26-0.26 & 1.34-1.47 \\
1999 Jan 11 & 14976s & 0.13-0.14 & 0.64-0.76 \\ 
1999 Jan 12 & 14080s & 0.27-0.36 & 1.79-3.00 \\ 
1999 March 6 & 4992s & 0.27-0.33 & 1.71-2.11 \\ 
1999 May 25 & 16000s & 0.32-0.47 & 2.36-3.25 \\
1999 May 26 & 16000s & 0.33-0.41 & 2.23-3.06 \\ 
1999 May 27 & 11520s & 0.18-0.20 & 0.94-1.02 \\

\enddata


\tablecomments{Col. 1: date. Col. 2: total integration
time in seconds. Col. 3: the range of optical depth
at the zenith at 850 $\mu$m for these observations. Col. 4:
same as for the previous column but at 450 $\mu$m.}

\end{deluxetable}

\clearpage
\begin{deluxetable}{crrrrr}
\footnotesize
\tablecaption{Catalog \label{tbl-2}}
\tablewidth{0pt}
\tablehead{
\colhead{Name} &
\colhead{Old name} & 
\colhead{RA and Dec (J2000.0)}   & 
\colhead{S/N} &
\colhead{$\rm S_{850 \mu m}$/mJy}   &
\colhead{$\rm S_{450 \mu m}$/mJy}
}
\startdata

CUDSS 14.1$^{\dag}$ & 14A &14 17 40.25 52 29 06.5 & 10.1 & 8.7$\pm$1.0 & 2.7$\pm$13.3\\
CUDSS 14.2$^{\dag}$ & 14B &14 17 51.7 52 30 30.5 & 6.3 & 5.5$\pm$0.9 & 22.9$\pm$12.0 \\
CUDSS 14.3$^{\dag}$ & ... &14 18 00.5 52 28 23.5 & 5.4 & 5.0$\pm$1.0 & -1.0$\pm$8.3 \\
CUDSS 14.4$^{\dag}$ & ... &14 17 43.35 52 28 14.5 & 5.3 & 4.9$\pm$0.9 & -9.8$\pm$15.7\\
CUDSS 14.5$^{\dag}$ & ... &14 18 07.65 52 28 21 & 4.5 & 4.6$\pm$1.0 & 12.6$\pm$7.5 \\
CUDSS 14.6 & ... &14 17 56.6 52 29 07 & 4.2 & 4.1$\pm$1.0 & -25.0$\pm$17.1\\
CUDSS 14.7$^{\dag}$ & ... &14 18 01.1 52 29 49 & 3.2 & 3.2$\pm$0.9 & 3.1$\pm$11.7\\
CUDSS 14.8$^{\dag}$ & 14E &14 18 02.7 52 30 15 & 4.0 & 3.4$\pm$0.9 & -5.5$\pm$11.2\\
CUDSS 14.9$^{\dag}$ & ... &14 18 09.0 52 28 04 & 4.1 & 4.3$\pm$1.0 & -2.6$\pm$7.5\\
CUDSS 14.10$^{\dag}$ & ... &14 18 03.9 52 29 38.5 & 3.5 & 3.0$\pm$0.8 & -6.4$\pm$11.7\\
CUDSS 14.11 & ... &14 17 47.1 52 32 38 & 3.5 & 4.5$\pm$1.3 & -12.7$\pm$18.5\\
CUDSS 14.12 & ... &14 18 05.3 52 28 55.5 & 3.4 & 3.4$\pm$1.0 & 7.4$\pm$8.6\\
CUDSS 14.13$^{\dag}$ & ... &14 17 41.2 52 28 25 & 3.4 & 3.3$\pm$1.0 & -1.4$\pm$15.2\\
CUDSS 14.14 & ... &14 18 08.65 52 31 03.5 & 3.3 & 4.6$\pm$1.3 & -22.0$\pm$15.2\\
CUDSS 14.15 & ... &14 17 29.3 52 28 19 & 3.1 & 4.8$\pm$1.5 & -22.3$\pm$25.2\\
CUDSS 14.16$^{\dag}$ & ... &14 18 12.25 52 29 20 & 3.7 & 4.7$\pm$1.4 & ......\\
CUDSS 14.17$^{\dag}$ & ... &14 17 25.45 52 30 44 & 3.3 & 6.0$\pm$2.1 & 35.7$\pm$41.3\\
CUDSS 14.18$^{\dag}$ & 14F &14 17 42.25 52 30 26.5 & 3.0 & 2.6$\pm$0.9 & 24.2$\pm$10.4\\
CUDSS 14.19$^{\dag}$ & ... &14 18 11.5 52 30 04 & 3.0 & 3.9$\pm$1.3 &4.4$\pm$18.8 \\

\enddata


\tablecomments{Col. 1: Name of source. A dagger indicates there
is a note about this source in the text. Col. 2: Name of
source in Paper I. Col. 3: Position of source. Col. 4:
Signal-to-noise with which the source
was detected at 850$\mu$m. Cols 5 \& 6: Flux density at 850 and
450 $\mu$m (measured using the methods described
in \S 4.1). The error does not include
the calibration error, which we estimate is $\simeq$5\% at
850 $\mu$m and $\simeq$20\% at 450 $\mu$m.}

\end{deluxetable}

\clearpage
\begin{deluxetable}{crrrrr}
\footnotesize
\tablecaption{Other Sources\label{tbl-3}}
\tablewidth{0pt}
\tablehead{
\colhead{Name} &
\colhead{Old name} &
\colhead{RA and Dec (J2000.0)}   &
\colhead{S/N} &
\colhead{$\rm S_{850 \mu m}$/mJy} &   
\colhead{$\rm S_{850 \mu m}$/mJy}    
}
\startdata

CUDSS 14.20 & 14D & 14 18 02.3  52 30 51.5 & 2.9 & 2.5$\pm$0.9 & 3.2$\pm$0.9 \\
CUDSS 14.21 & 14C & 14 17 33.8  52 30 49 & 2.4 & 2.3$\pm$1.0 & 3.8$\pm$1.1\\

\enddata
 
 
\tablecomments{Col. 1: Name of source. Col. 2: Name of
source in Paper I. Col. 3: Position of source. Col. 4:
Signal-to-noise with which the source
was detected at 850$\mu$m. Col. 5: Flux density at 850$\mu$m.
Col. 6: Flux density at 850$\mu$m from Paper I.}
 
\end{deluxetable}


\clearpage

\begin{deluxetable}{crrrrr}
\footnotesize
\tablecaption{SCUBA/ISO Associations\label{tbl-4}}
\tablewidth{0pt}
\tablehead{
\colhead{CUDSS} &
\colhead{ISO} &
\colhead{CFRS} &
\colhead{RA and Dec (J2000.0)}   & 
\colhead{d}   &
\colhead{p}
}
\startdata

14.13 & 0 & 14.1157 & 14 17 41.81 52 28 23.0 & 5.9 & 0.03 \\
14.17 & 195 & 14.1569 & 14 17 24.36 52 30 46.45 & 10.3 & 0.08 \\ 
14.18 & 5 & 14.1139 & 14 17 42.04 52 30 25.7 & 2.1 & 0.0036 \\

\enddata


\tablecomments{Col. 1: Name of SCUBA source. Col. 2: Name of ISO
source using the nomenclature of Flores et al. (1999b). Col. 3:
Name of galaxy associated with the ISO source using the standard
CFRS nomenclature (Flores et al. 1999b). Col. 4: Position of the
galaxy in J2000 coordinates. Col. 5: Distance in arcsec between SCUBA
position and optical position. Col. 6: Probability of an unrelated
ISO source
falling closer to the SCUBA source than this distance.}

\end{deluxetable}

\clearpage

\begin{deluxetable}{crrrrrrrr}
\footnotesize
\tablecaption{SCUBA/Radio Associations\label{tbl-5}}
\tablewidth{0pt}
\tablehead{
\colhead{CUDSS} &
\colhead{Radio} &
\colhead{RA and Dec (J2000.0)}   & 
\colhead{d}   &
\colhead{p} &
\colhead{$\rm S_{5 GHz}$} &
\colhead{$\alpha$} &
\colhead{$z_{est}$} &
\colhead{$z_{act}$}} 

\startdata

14.1 & 15V18 & 14 17 40.32 52 29 05.9 & 0.9 & $\rm 6.0 \times 10^{-4}$ & 44.0 & 
0.2 & $\rm 2.2\pm0.5$ & ... \\
14.2 & .... & ................... & ... & ... & $<$19 & ... & $>$1.95 & ... \\
14.3 & 15V53 & 14 18 00.5 52 28 20.8 & 2.7 & $\rm 8.6 \times 10^{-3}$ & 30.1 & 0.9 & 
$\rm 1.3\pm0.3$ & ... \\
14.4 & .... & ................... & ... & ... & $<$15 & ... & $>$2.1 & ...  \\
14.5 & .... & ................... & ... & ... & $<$12 & ... & $>$2.2 & ...  \\
14.6 & .... & ................... & ... & ... & $<$12 & ... & $>$2.1 & ... \\
14.7 & .... & ................... & ... & ... & $<$15 & ... & $>$1.7 & ... \\
14.8 & .... & ................... & ... & ... & $<$18 & ... & $>$1.6 & ... \\
14.9 & 15V67 & 14 18 09.03 52 28 03 & 1.0 & $\rm 1.1 \times 10^{-3} $ & 46.1 & -0.2 & $\rm 2.0\pm0.5$ & ...  \\
14.10 & .... & ................... & ... & ... & $<$18 & ... & $>$1.5 & ... \\
14.11 & .... & ................... & ... & .... & ($<$83) & ... & $>$1.3 & ...\\
14.12 & .... & ................... & ... & ... & $<$12 & ... & $>$1.9 & ...\\
14.13 & .... & 14 17 40.67 52 28 24.6 & 4.9 & $\rm 2.1 \times 10^{-2} $ & 15 & ... & $\rm 2.1\pm0.5$ & ... \\
            & 15V23 & 14 17 41.81 52 28 23.4 & 5.8 & $\rm 2.9 \times 10^{-3}$ & 53.6 & 0.3 & $\rm 1.2\pm0.3$ & 1.15 \\
14.14 & .... & ................... & ... & ... & $<$26 & ... & $>$1.5 & ...\\
14.15 & .... & ................... & ... & .... & ($<$75) & ... & $>$1.4 & ...\\
14.16 & .... & ................... & ... & ... & $<$15 & ...& $>$2.0 & ...\\
14.17 & .... & ................... & ... & .... & ($<$49) & ...  & $>$2.0 & ...\\
14.18 & 15V24 & 14 17 42.08 52 30 25.2 & 2.0 & $\rm 2.4 \times 10^{-4}$ & 78.9 & 0.5 & $\rm 0.7\pm0.2$ & 0.66 \\
14.19 & .... & ................... & ... & ... & $<$16 & ... & $>$1.8 & ...\\

\enddata


\tablecomments{Col. 1: Name of SCUBA source.
Col. 2: Name of radio source \cite{fom}. Col. 3: Position of radio
source.
Col. 4: Distance in arcsec between SCUBA and
radio positions. Col. 5: Probability of a radio source lying within
this distance of the SCUBA source by chance. Col. 6: 
Flux density at 5 GHz in $\mu$Jy. CUDSS 14.11, 14.15 and 14.17 are
outside the useable region of the 5-GHz map, so for these sources
the radio upper limits are the 3$\sigma$ upper limits obtained
from the re-reduced 1.5GHz map (see text).
Col. 7: Radio spectral index, $\alpha$, from Fomalont et al.
(1991), defined such that $\rm flux \propto frequency^{-\alpha}$.
Col. 8: Redshift estimated from the ratio of 850$\mu$m 
to radio flux (see text). Col. 9: Spectroscopic redshift.
}

\end{deluxetable}

\clearpage
\begin{deluxetable}{crrrrrr}
\footnotesize
\tablecaption{Redshift Distributions\label{tbl-6}}
\tablewidth{0pt}
\tablehead{
\colhead{Sample} &
\colhead{sources} &
\colhead{$z_{spect}$} &
\colhead{$z_{phot}$} &
\colhead{Mean $z$} &
\colhead{Median $z$} &
\colhead{95\% limits}}
\startdata
 
CUDSS 14$^h$& 19 & 2 & 3 & 1.96$\pm$0.15 & 2.05 & undefined \\
Clusters& 16 & 4 & 5 & 2.70$\pm$0.39 & 2.52 & 1.59,3.43 \\
HFF& 7 & 0 & 5 & 2.51$\pm$0.81 & 1.5 & 1.15,4.0 \\
combined& 42 & 6 & 13 & 2.81$\pm$0.36 & 2.41 & 1.94,3.48 \\
 
\enddata
 
 
\tablecomments{Col. 1: Name of sample with the following key:
CUDSS 14$^h$---this paper; Clusters---sample from Smail et al.
(2000); HFF--sample from Barger, Cowie \& Richards (2000); combined---all
of the above samples together. Col. 2: Number of sources in the sample.
Col. 3: Number of sources with spectroscopic redshifts; Col. 4:
Number of sources with photometric redshifts but not
a spectroscopic redshift. Col. 5: 
Mean redshift and error estimated
using the Kaplan-Meier estimator; Col. 6---Median redshift; Col. 7---95\% 
confidence
interval for the median.}

\end{deluxetable}
 
\clearpage

\clearpage





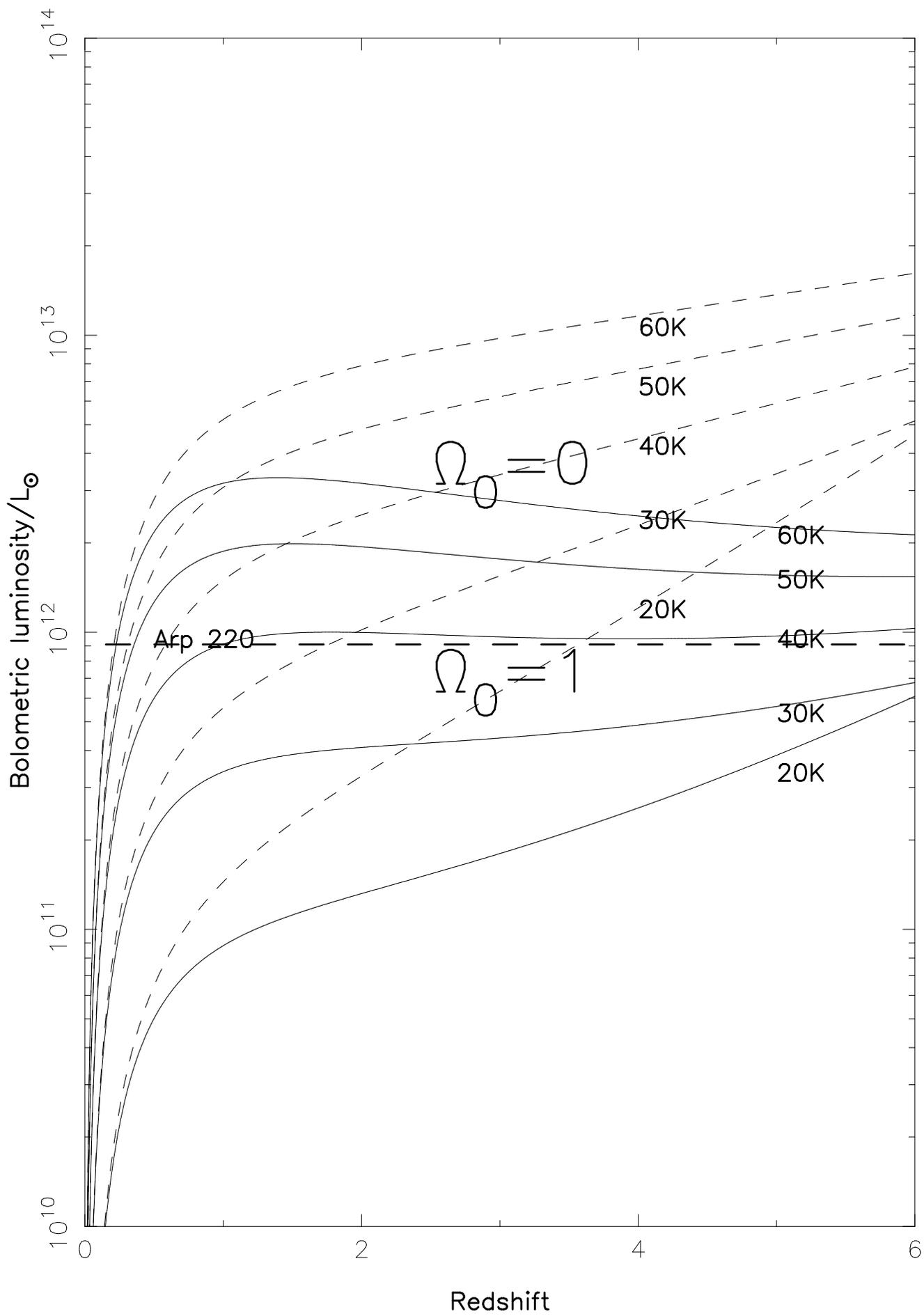
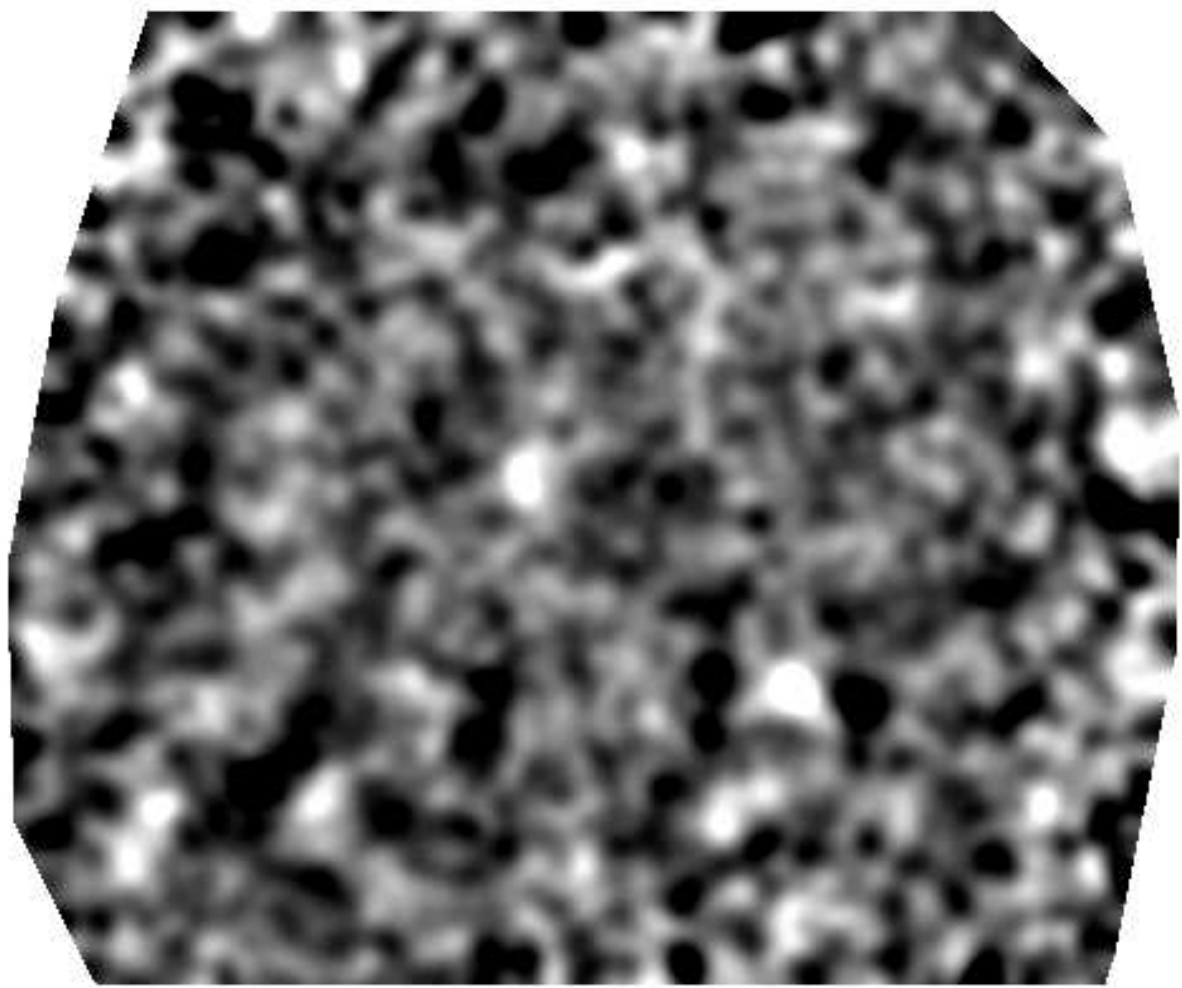
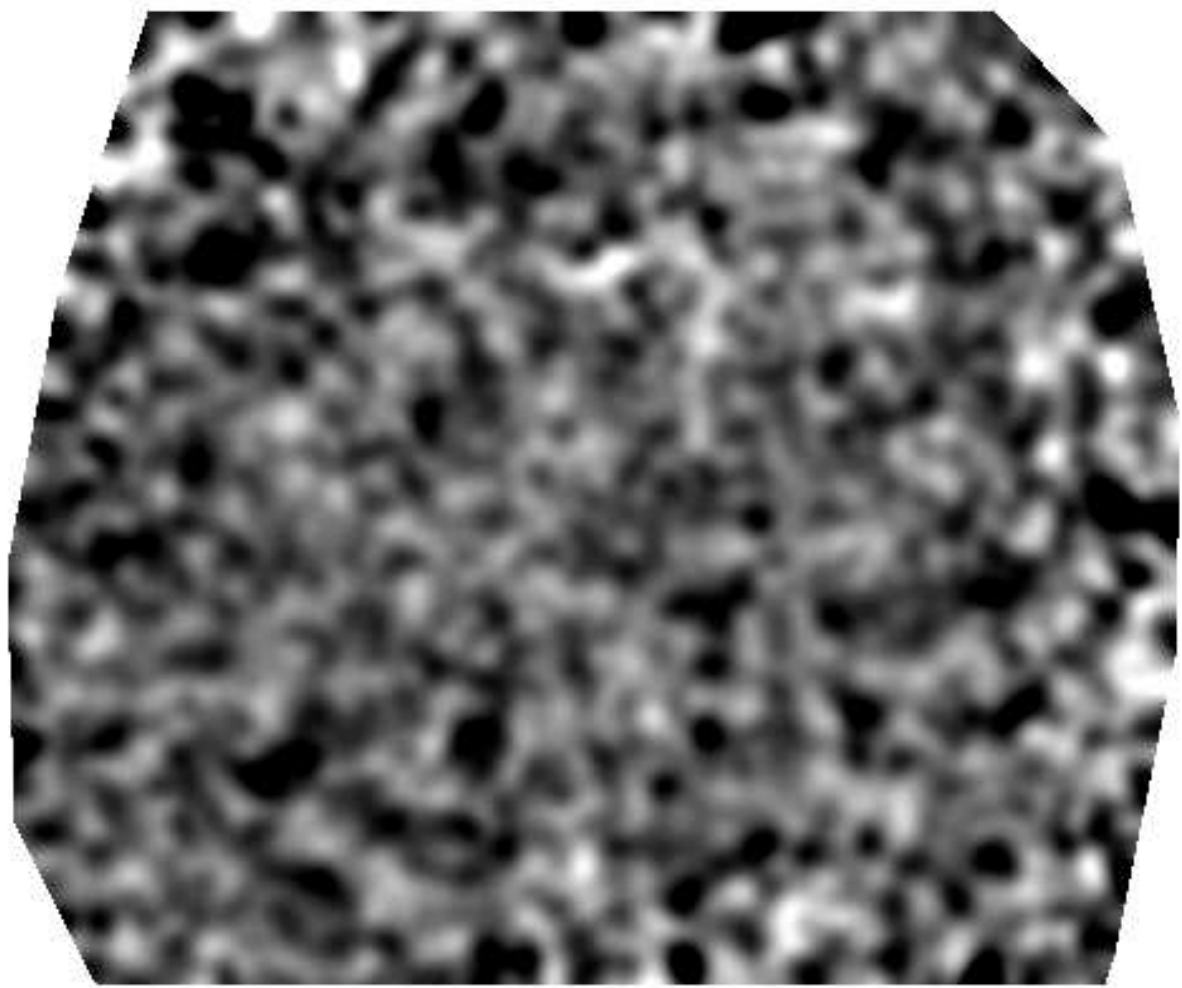
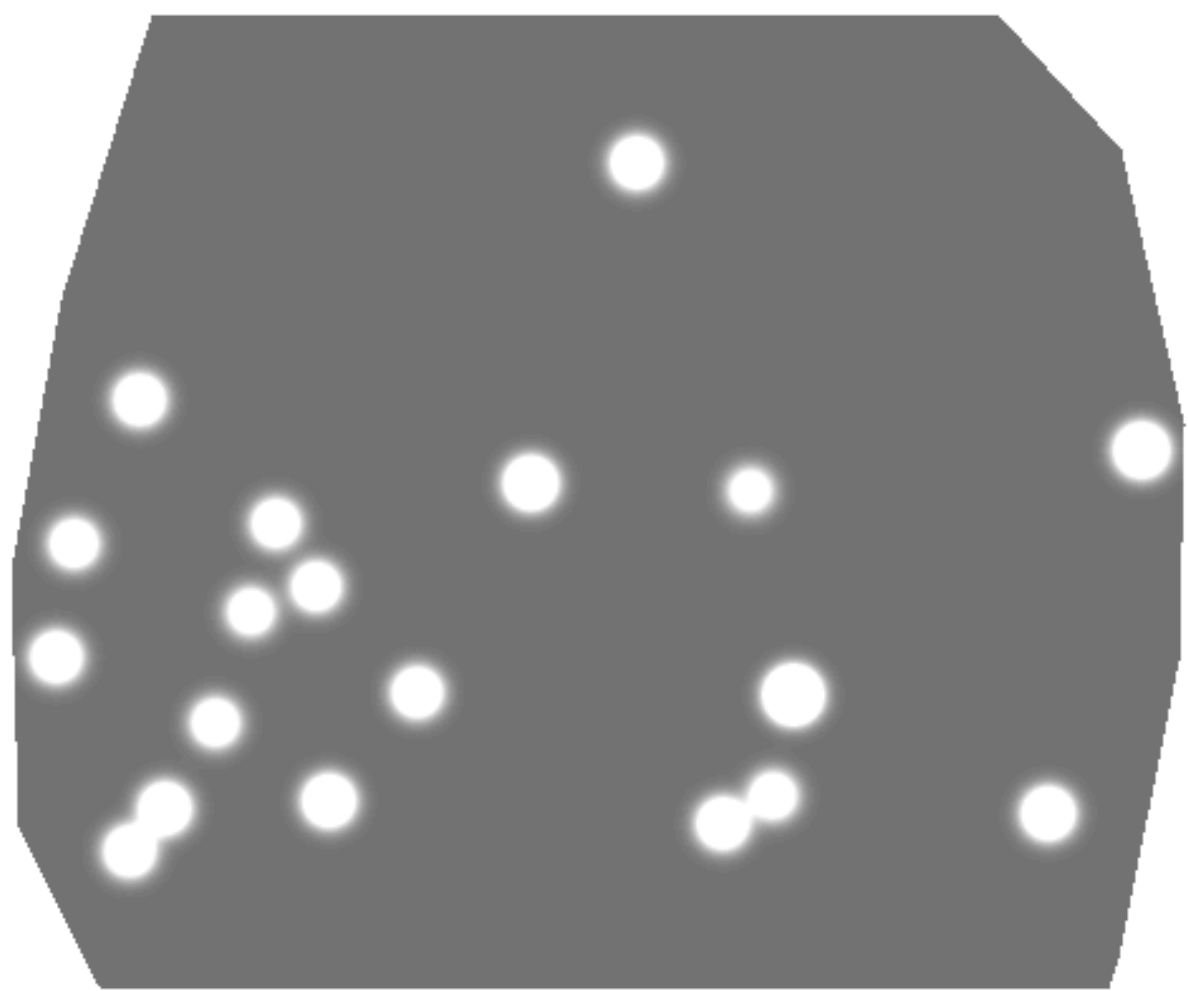
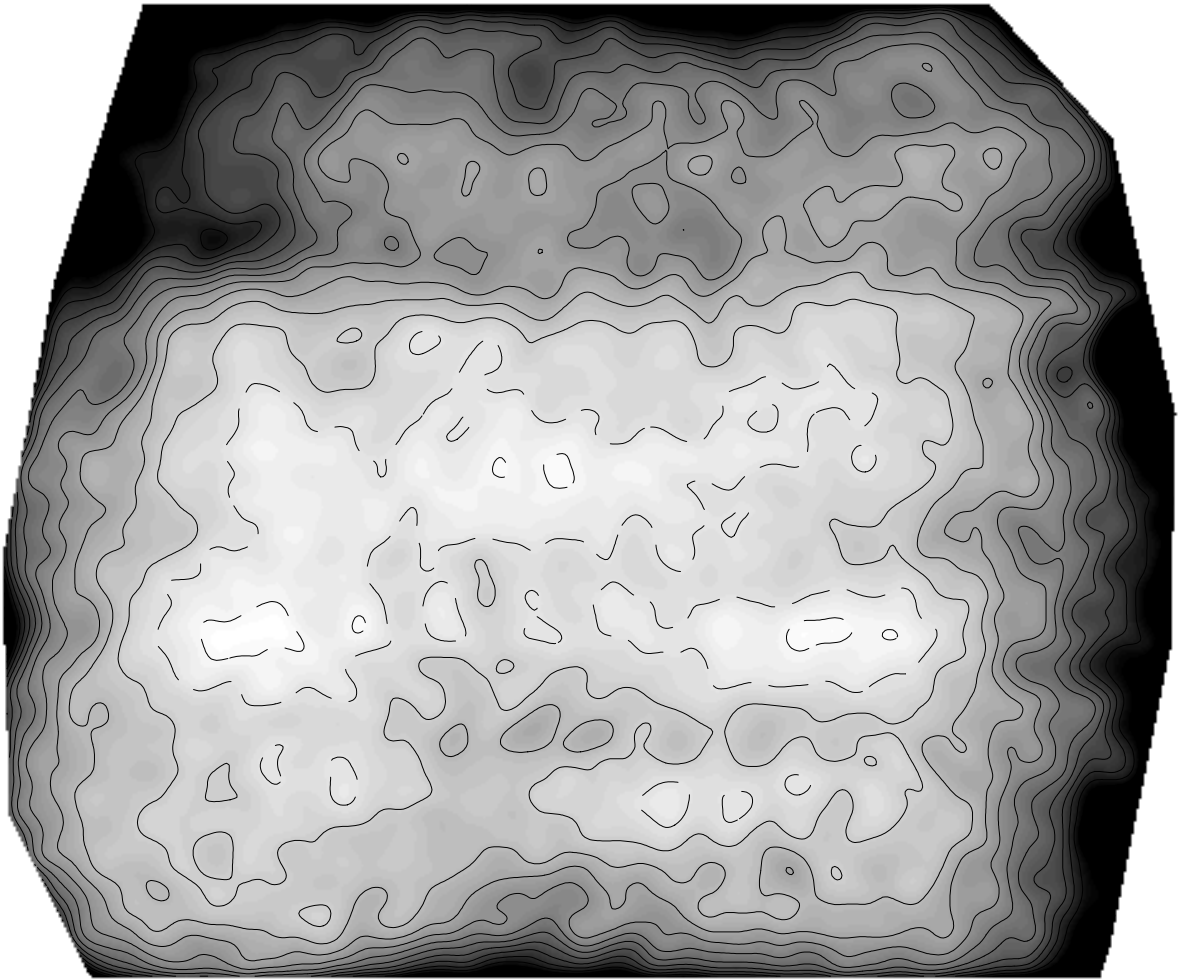
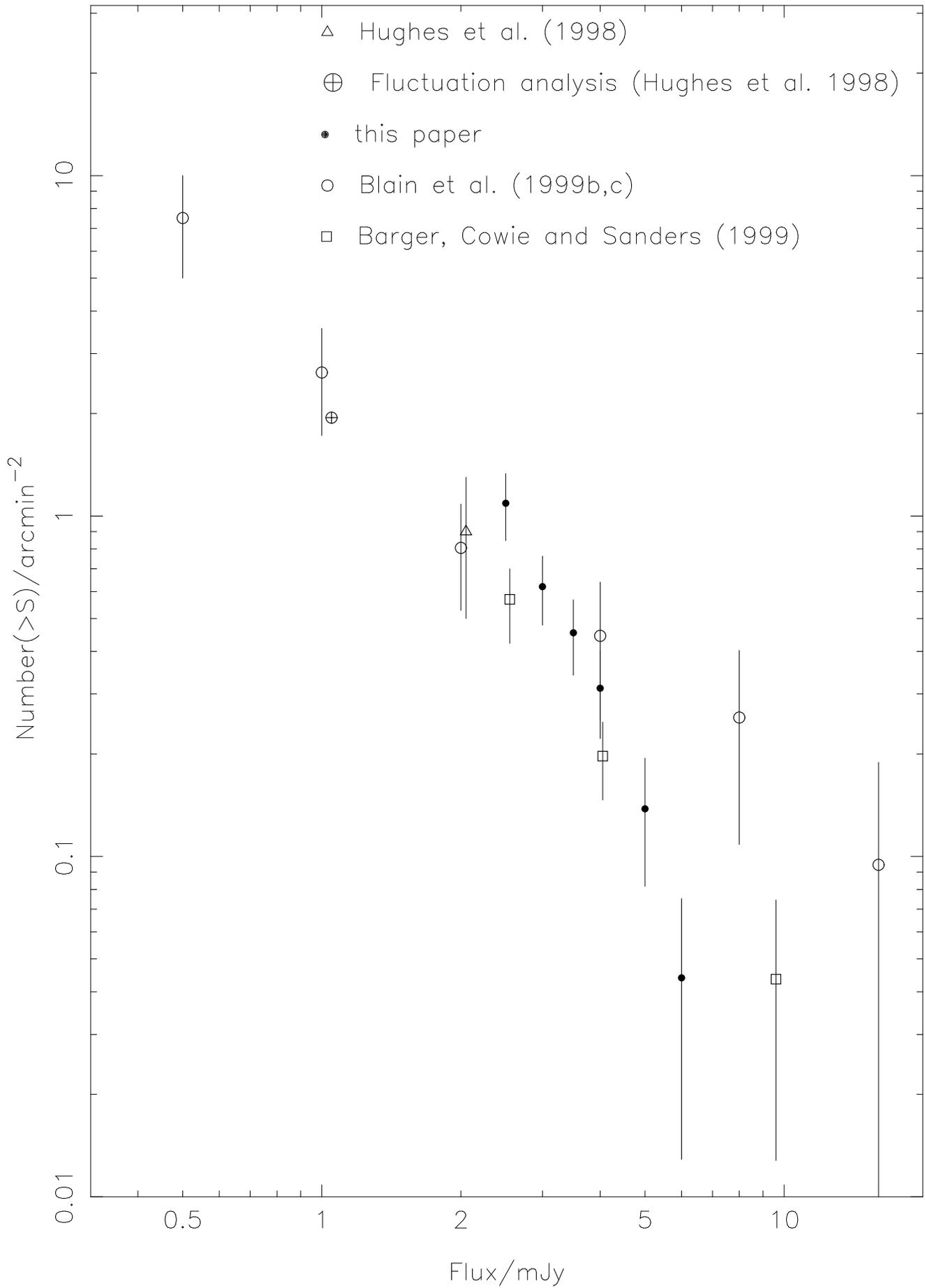
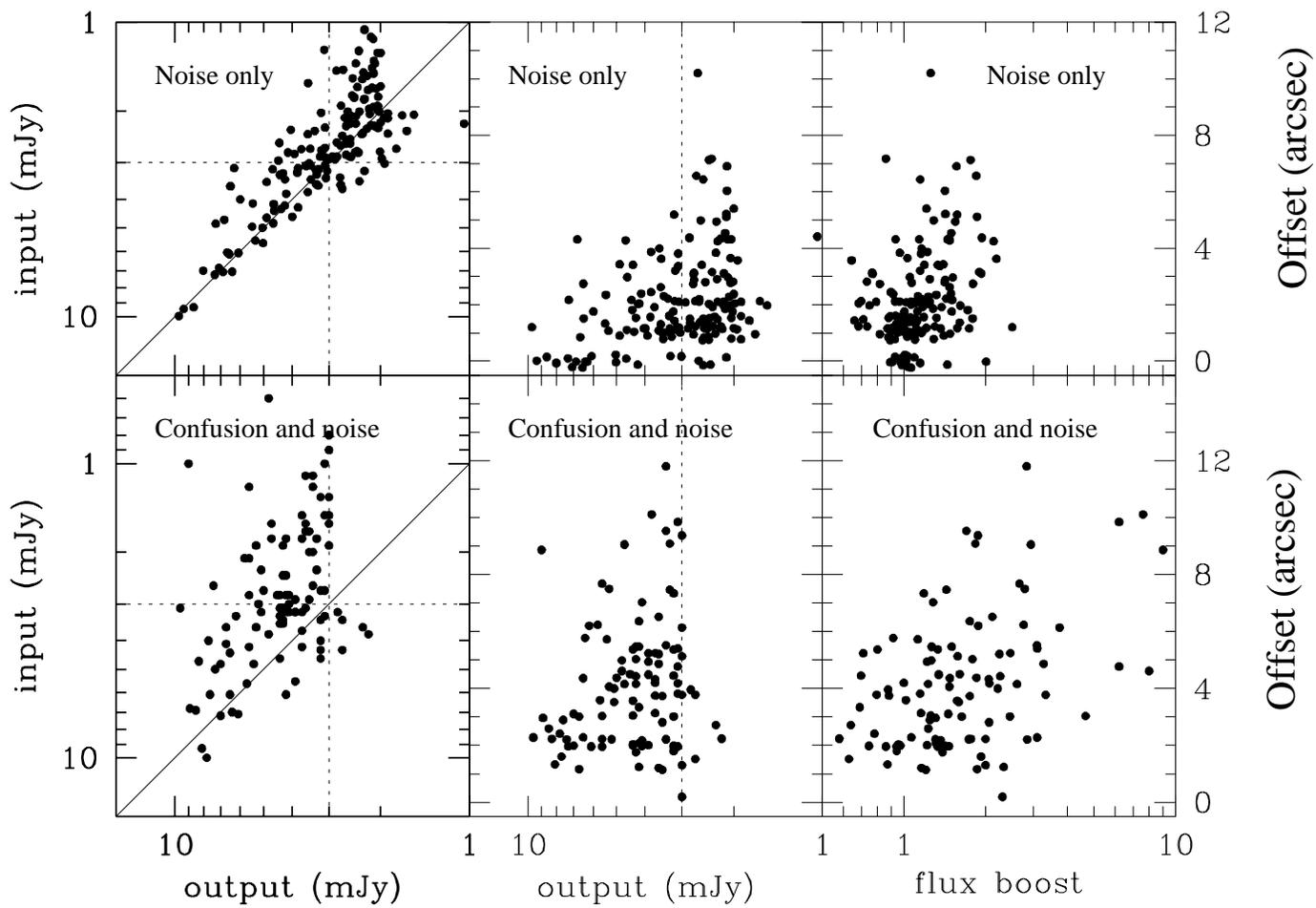
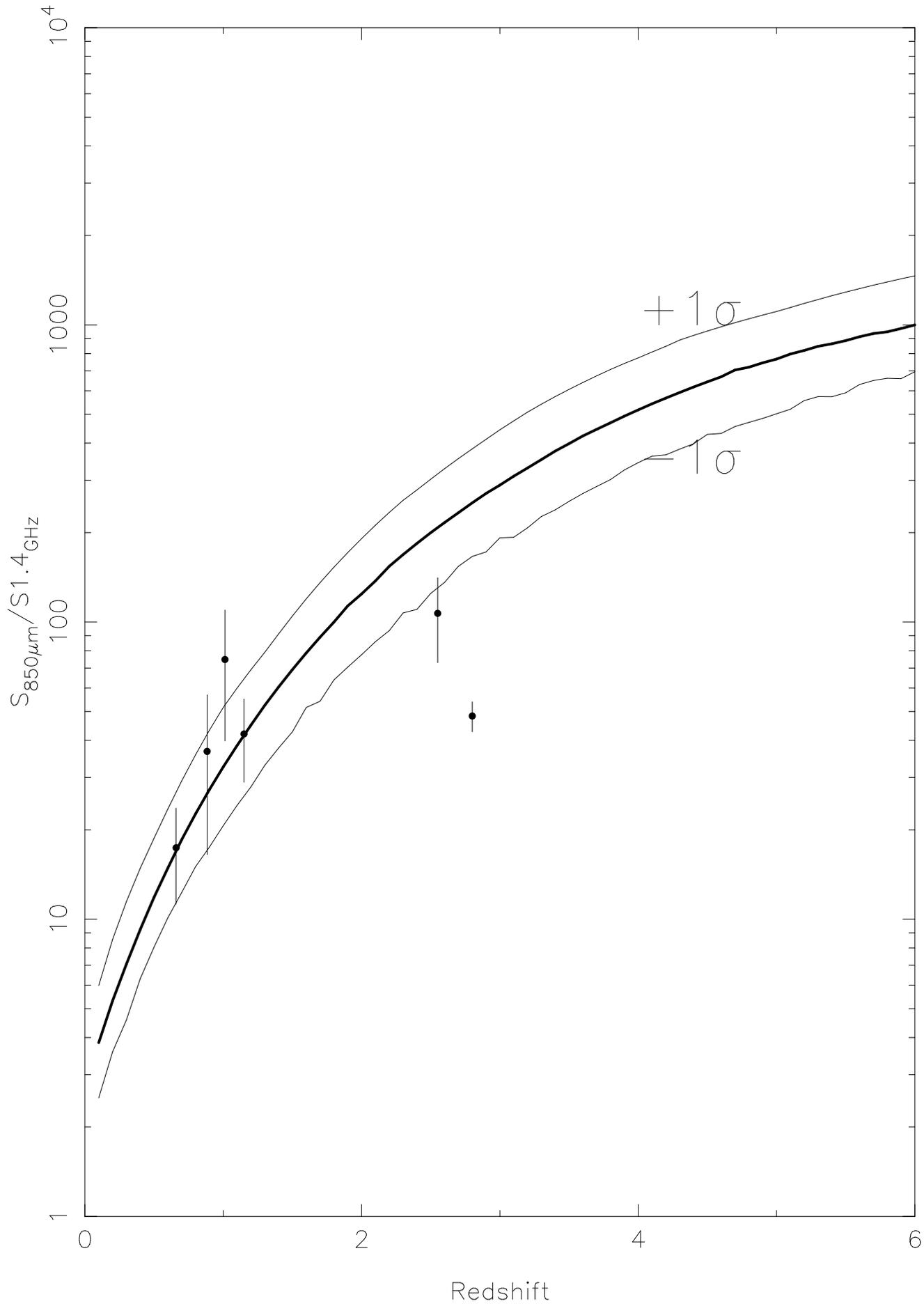
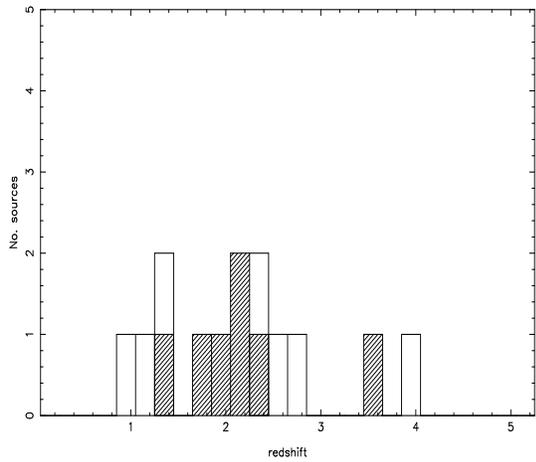
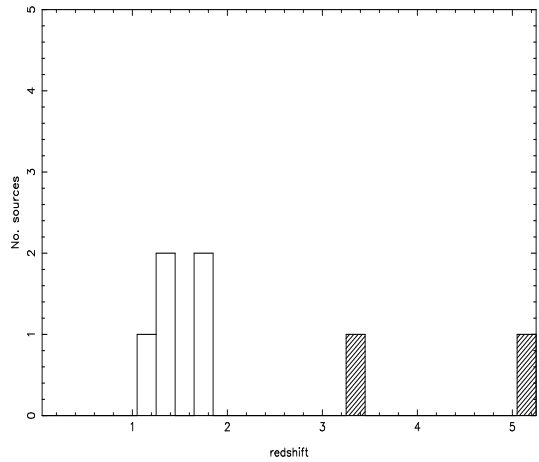
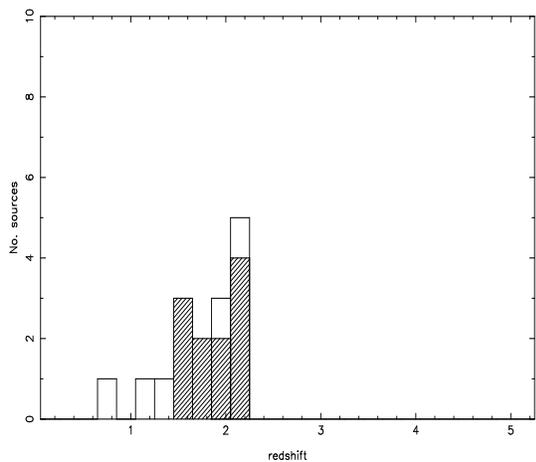
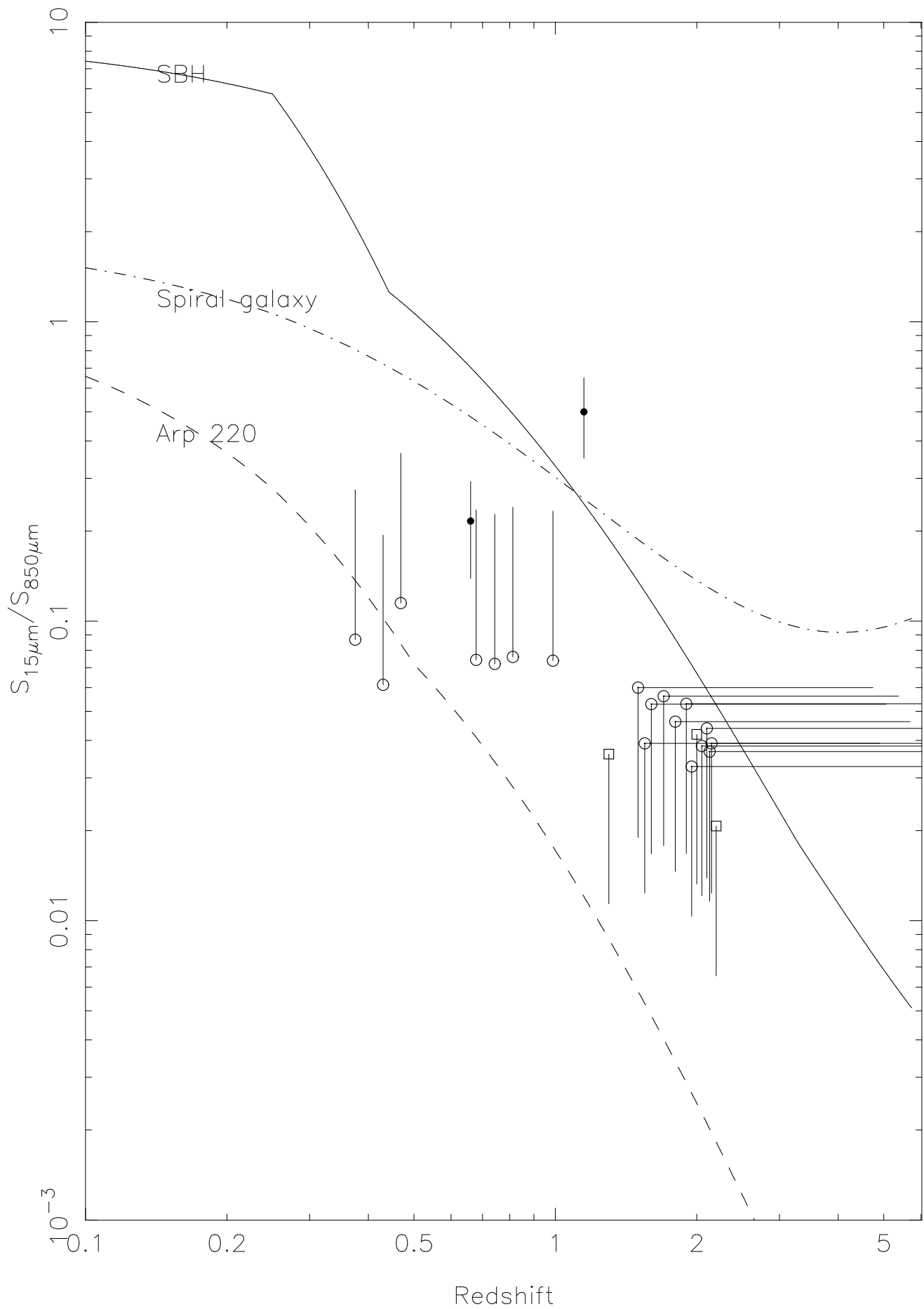
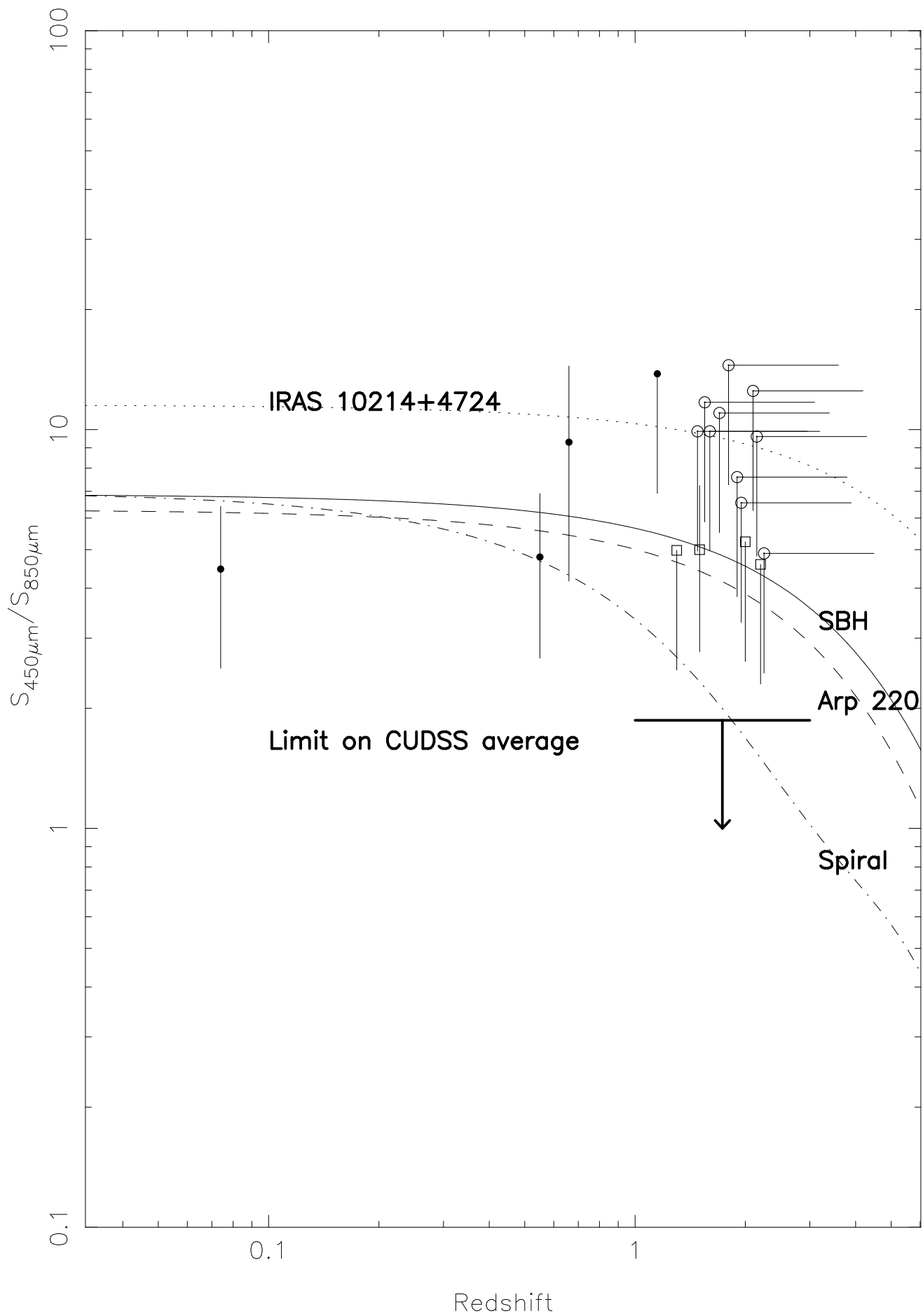
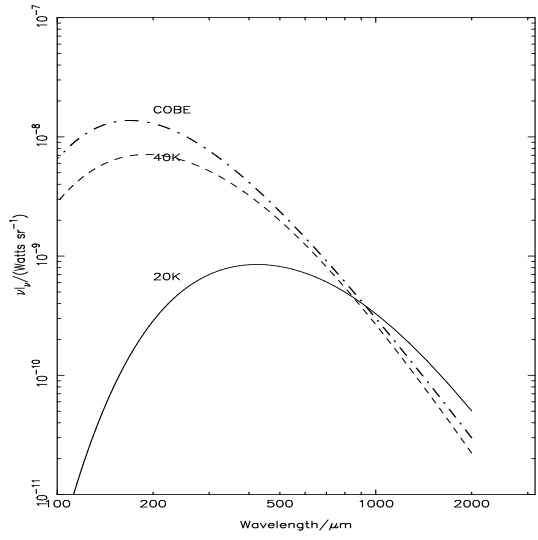
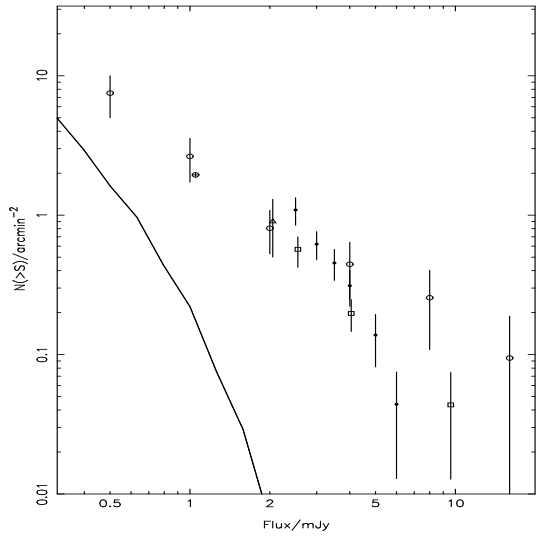
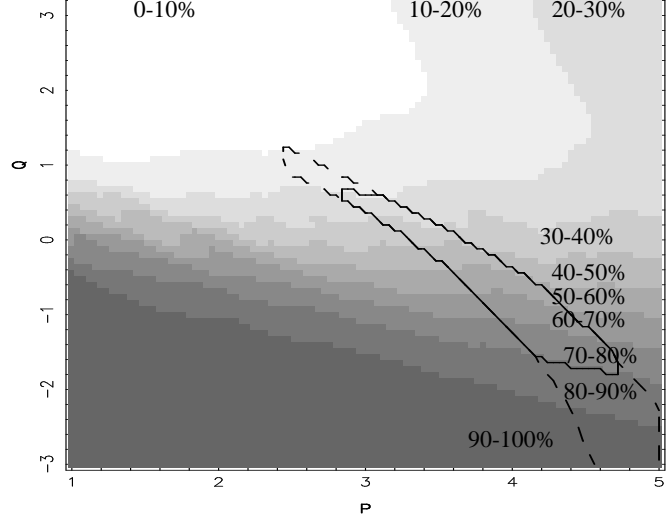
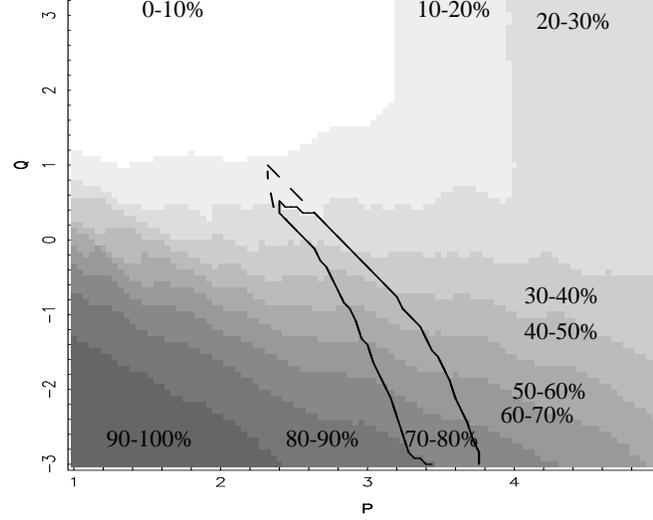
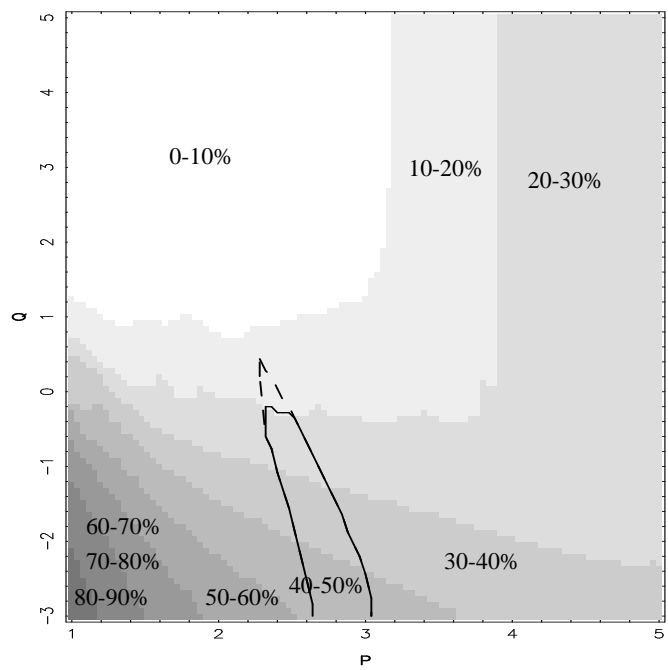
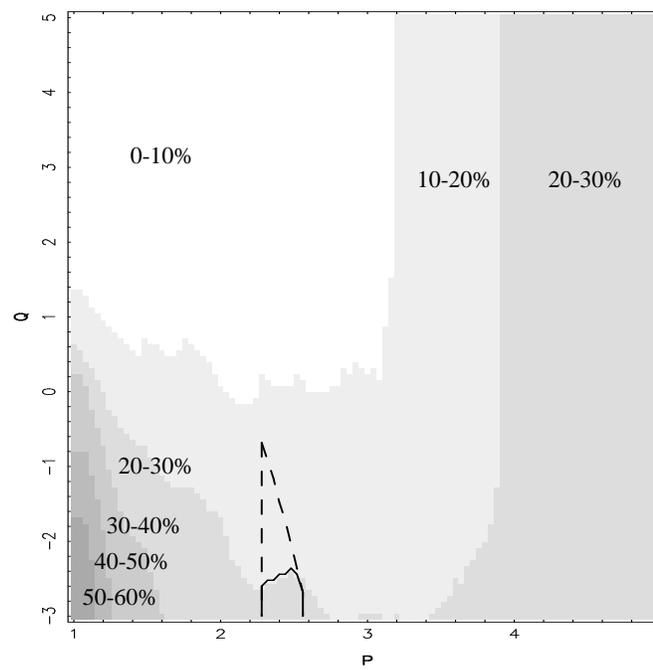
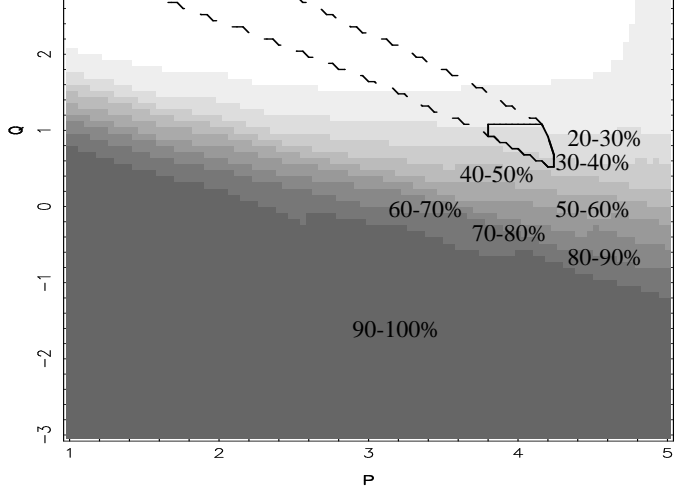
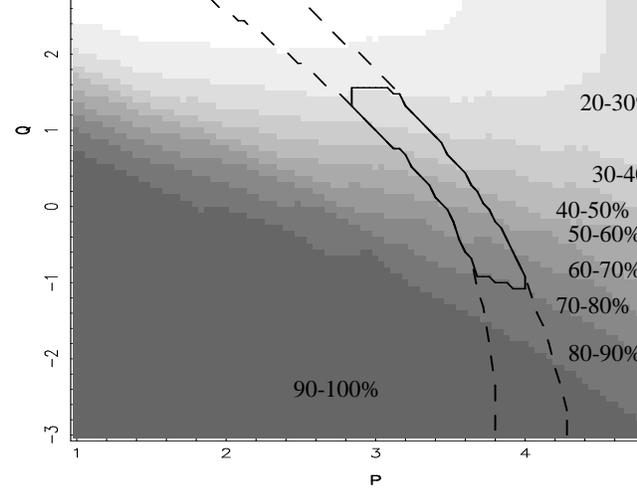
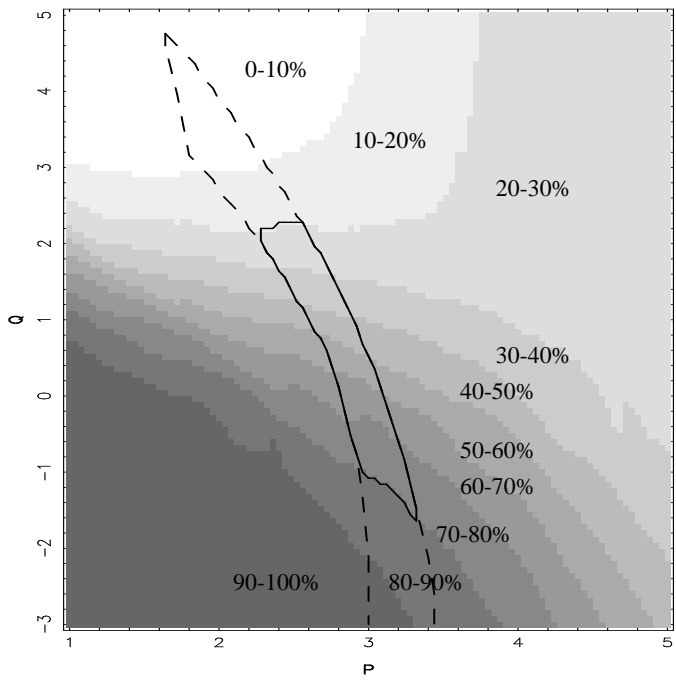
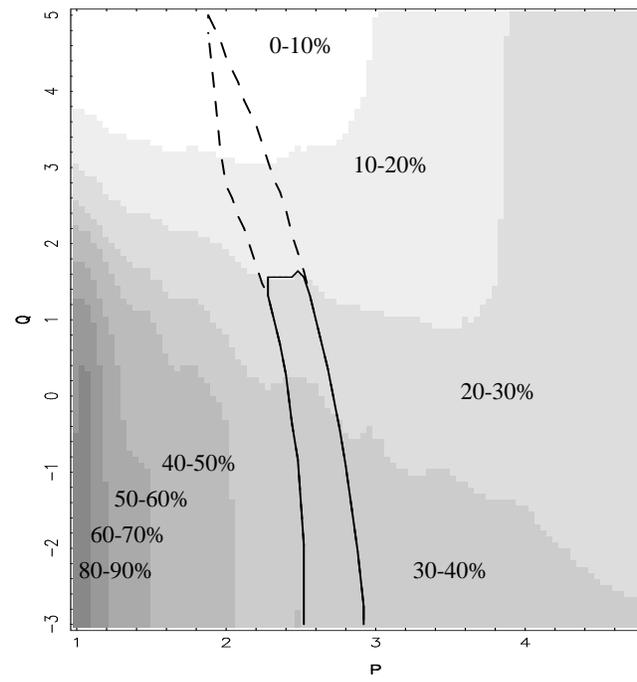
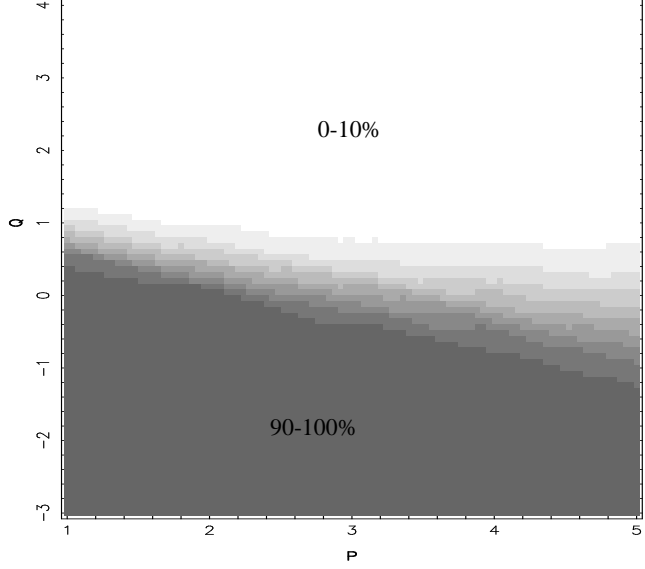
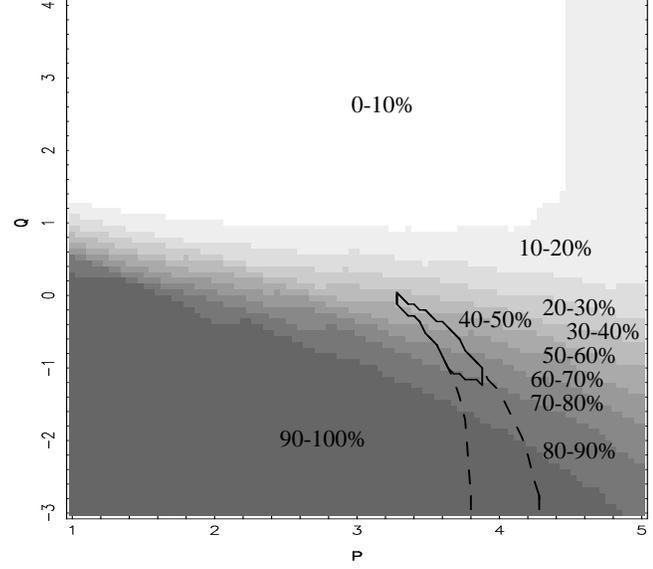
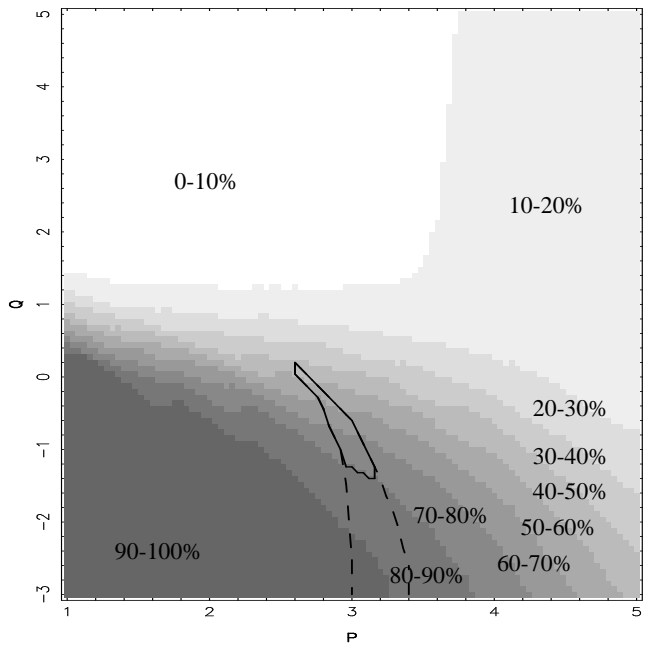
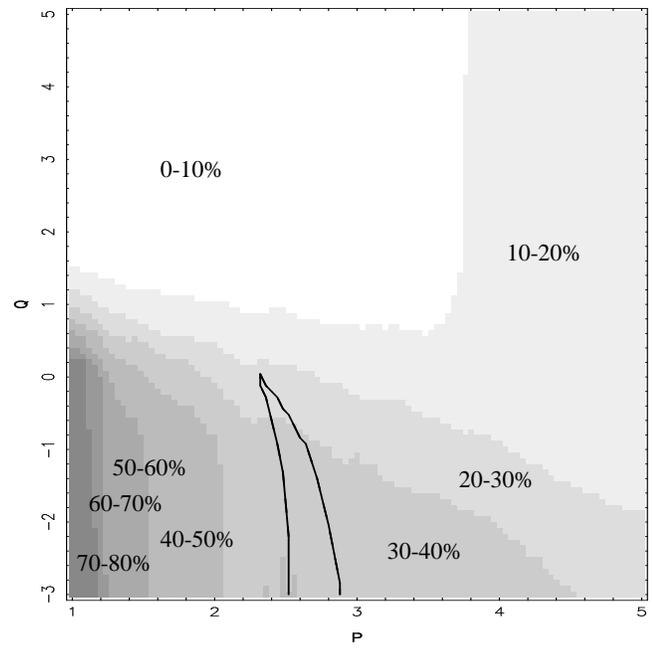
\end{document}